\documentclass[11pt]{article}
\usepackage[T1]{fontenc}        
\usepackage{lmodern}            
\usepackage[american]{babel}
\usepackage[dvips]{graphicx,color} 
\usepackage{amsmath,amssymb}
\usepackage{mathtools}
\newcommand{\bce}{\begin{center}}
\newcommand{\ece}{\end{center}}
\newcommand{\bea}{\begin{eqnarray}}
\newcommand{\eea}{\end{eqnarray}}
\newcommand{\be}{\begin{equation}}
\newcommand{\ee}{\end{equation}}
\newcommand{\bd}{\begin{displaymath}}
\newcommand{\ed}{\end{displaymath}}
\newcommand{\bit}{\begin{itemize}}
\newcommand{\eit}{\end{itemize}}
\newcommand{\ben}{\begin{enumerate}}
\newcommand{\een}{\end{enumerate}}
\newcommand{\bdes}{\begin{description}}
\newcommand{\edes}{\end{description}}

\newcommand{\E}{\> = \>}

\newcommand{\non}{\nonumber\\}
\newcommand{\To}{\> \longrightarrow \>}
\newcommand{\Def}{\> \coloneqq \>}

\newcommand{\Id}{\textnormal{ \raisebox{-0.4pt}{{\large {\bf 1}}}}}
\newcommand{\drl}{\, \overleftrightarrow{\partial} \,}  
\newcommand{\la}{\left\langle \,}
\newcommand{\ra}{\, \right\rangle}
\newcommand{\lrp}{\left ( \, }    
\newcommand{\rrp}{\, \right ) }   
\newcommand{\lsp}{\left [ \, }    
\newcommand{\rsp}{\, \right ] }   
\newcommand{\lcp}{\left \{ \, }   
\newcommand{\rcp}{\, \right \} }  
\def\lvl{\, \left | \, }     
\def\rvl{\, \right | \, }    
\title{
  {\vspace{-2cm} \normalsize
     \includegraphics[width=80mm]{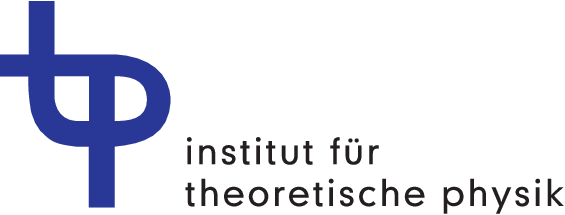}
     \hfill\parbox[b][30mm][t]{35mm}{MS-TP-22-02}%
  }\\[25mm]
Scalars from Gauge Fields
      }
\author{M.~Stingl%
        \thanks{Retired. E-mail address: stingl@uni-muenster.de}  \\
        Institut f\"ur Theoretische Physik,
        Universit\"at M\"unster                                   \\
        Wilhelm-Klemm-Str.~9, D-48149 M\"unster, Germany}       
\date{January 25, 2022}
\begin{document}
\maketitle
\begin{abstract}
\noindent
In an Euclidean SU(2) $\otimes$ U(1) gauge theory without fermions, we
identify scalar-field variables, functionals of the gauge fields and coming
in different representations of isospin, which (i) are of mass dimension
one in $d=4$, (ii) couple to their parent gauge fields through suitable
gauge-covariant derivatives, and (iii) can be endowed with a hypercharge
despite their parents having none. They can be interpreted as projections
of the gauge vectors onto an orthonormal basis that is defined by the fields
themselves. We inquire as to whether these scalars can perform the usual
tasks, normally fulfilled by external scalar fields, of spontaneous symmetry
breaking and mass generation through vacuum expectation values. The gauge
Lagrangian, expressed in terms of these scalars, automatically has quartic
and cubic terms; no extra coupling constant for quartic scalar
self-interactions is needed. VEV formation takes place in one of four scalar
fields populating the classical potential-energy minimum. There are nine
massive Higgs particles, a neutral triplet at a mass of $m_Z \sqrt{2}$, and
three conjugate pairs of charged ones at $m_W \sqrt{2}$. Seven quasi-Goldstone
scalars remain massless. This results in a qualitatively correct pattern of 
heavy-vector masses and mixing, with the analog of the mixing angle determined
by theory. Higgs-type hypercharge and charge assignments emerge naturally.
\end{abstract}
\newpage


\section{In search of ``intrinsic'' scalars}
\setcounter{equation}{0}
The Lorentz-scalar Higgs fields providing for spontaneous symmetry breaking 
and mass generation in present-day electroweak theory are 
introduced {\em in addition} to the gauge-vector fields. In this article
we refer to those as scalars {\em external} to the gauge-vector system.
It is a valid question whether such fields could, alternatively, 
arise from the dynamics of the gauge fields themselves. The
latter, after all, possess self-interactions that could in principle stabilize
scalar combinations with distinguishable experimental signals. Clearly, such
{\em intrinsic} scalars, if indeed they are to play their assigned roles
in electroweak theory, would from the outset have to meet nontrivial
restrictions. To make mass formation along the known lines feasible, they
would, in addition to being true Lorentz scalars, have to couple to their
parent gauge fields through something like gauge-covariant derivatives. For
this, in turn, they would need to be, like ordinary scalar fields, of mass
dimension one in $d=4$ dimensions (or $1 - \epsilon$ in $d=4-2\epsilon$
dimensions) in order for these interactions to fit into, or be derivable
from, the known renormalizable gauge-field action. They would, moreover,
have to carry the extra property of hypercharge, since it is only through
this quantum number that the abelian U(1) field can couple dynamically.
It is not clear a priori whether all these properties can be realized
simultaneously. The question is not merely one of economy: it is known that
the quartic self-interaction of the ``external'' scalar fields does not
allow for asymptotic freedom, and it is still true that
``there is today a widespread feeling that interacting quantum fields 
that are not asymptotically free .... are not mathematically 
consistent''\cite{Wein}. In the following we hope to elucidate some sides 
of the above questions, not least because their study may reveal aspects 
of gauge theory itself that do not seem to have attracted attention previously. 

       In what follows, section 2 initiates, and section 3 completes, the
identification of  ``intrinsic'' scalar fields with desirable properties
as functionals of the gauge-vector fields. This is done for the simplest
case of a pure U(2) gauge theory, without either charged-vs.-neutrals
mixing or coupling to fermions. Without the introduction of hypercharge,
this theory is partly trivial, since in the absence of both fermions and
external scalars, and without neutral-sector mixing, the abelian field
decouples.  It will however provide a simple workspace in which to try
things out.

      Section 4 leads, in several steps, to the identification of these
scalars with projections of the gauge vectors onto an orthonormal basis
defined by the vector fields themselves. Section 5 discusses the rewriting
of the gauge action in terms of ``intrinsic'' scalars, for the case where
the latter are not yet endowed with
hypercharge. Still for the same case, section 6 describes the modification
of the usual concept of a covariant derivative that becomes necessary in
building gauge or BRS-invariant terms for the action. Section 7 suggests
a way of conferring a hypercharge onto the scalars, despite the fact that
their parent gauge fields carry none, and of adapting the covariant-derivative
concept to this situation. On this basis, section 8 then discusses a ``mixed''
formulation of the action, where {\em both} types of fields, and their
relationship as enforced through suitable delta functionals, make their
appearance. Section 9 analyzes the modified, gauge-covariant derivative in
more depth and dicusses its transformation to an isospin representation
making electric charge diagonal, such that electrically neutral scalars,
capable of formation of vacuum expectation values, can be identified. Along
the way, Higgs-type charge and hypercharge assignments emerge naturally.

      In this framework, section 10 finally takes up the issue of mass
formation, both for the heavy vector bosons and for the scalars
themselves. It will be seen that a qualitatively correct, and even
semi-quantitative, pattern of vector masses and mixings results without
further assumptions. In addition there are a set of nine massive
Higgs-type scalars, both charged and neutral, at a fixed mass ratio to their
vector counterparts. In closing, section 11 offers an enumeration of
questions not yet taken up or unresolved.

      That one and the same action functional, when written in different
variables, should describe different excitations of the same underlying
field system is not, of course, a new phenomenon. A well-known example is
Coleman's rewriting\cite{Cole} of the massive Thirring model in terms
of sine-Gordon solitons, where the solitonic excitation appears as a kind
of coherent state built from the ''particle'' excitations. The analogy is not
as farfetched as it would seem: we shall see in sect.~3 that the ``intrinsic''
scalars contain factors that can indeed be read as coherent superpositions
of gauge-vector excitations.
\vspace{0.5cm}

\section{Introducing the Scalar Fields}
\setcounter{equation}{0}

         The many {\em composite} scalar fields that may be formed from the 
set of four vector fields of an  SU(2) $\otimes$ U(1) theory clearly are
not the answer to our needs, as they all are of higher mass dimension. 
Although we will begin with a simple field of this type -- one
that is easily identifiable in the standard gauge Lagrangian -- we will then
have to perform a process of ``extraction of roots'', in a sense to be made
precise below, to arrive at fields of mass dimension one.    
                                    
       We denote, as usual, by $A_{\mu}^{a} \, \lrp x \rrp$ the three SU(2)
fields with isospin indices $ a \E 1, 2, 3$, and by $B_{\mu} \lrp x \rrp$ the
U(1) field in a four-dimensional Euclidean $x$ space, but we shall soon be
in need of a uniform notation for all four fields and thus write 
\be
A_{\mu}^{C} \, \lrp x \rrp , \  C \E 1, 2, 3, 4, \qquad \mathrm{with} \  
             A_{\mu}^{4} \, \lrp x \rrp  \Def B_{\mu} \, \lrp x \rrp \  ,
\label{Afour}
\ee
with a capital index running over the four directions of an extended isospace
(technically, the adjoint-representation space of the group U(2)),
while lower-case indices will continue to take their three values, and
appropriate summation conventions apply to both. (Thus, for example,
$\delta^{A\,c}\,\delta^{B\,c}\ +\ \delta^{A\,4}\,\delta^{B\,4} 
\E \delta^{A\,B}$).
Then in the standard (Euclidean) gauge-field action
\be
S_E \lsp A,\bar c, c \rsp  \E  S_{2}\lsp A \rsp \,+\, S_{3}\lsp A \rsp \,+\,
                         S_{4}\lsp A \rsp \,+\, S_{GFG}\lsp A, c,\bar{c} \rsp ,
\label{gaugact}
\ee
where $S_{GFG}$ contains gauge-fixing-and-ghosts terms, and where $S_{(2\,,3\,,
4)}$ denote the bilinear, trilinear, and quadrilinear portions of the 
classical action, the bilinear term might be written as
\be
S_{2}\lsp A \rsp  \E  \frac{1}{4} \, \int \! d^4 x \, \lrp \partial_{\mu}
A_{\nu}^{C} \, - \, \partial_{\nu} A_{\mu}^{C} \rrp ^2   \E   \frac{1}{2} \, 
\int \! d^4 x \, A_{\mu}^C \lsp \delta_{\,\mu \,\nu}\,(-\partial_{\lambda}
\partial_{\lambda} )\, + \, \partial_{\mu} \,\partial_{\nu} \rsp A_{\nu}^C,
\label{S2}
\ee
whereas the $S_{3,4}$ terms would involve lower-case isospin indices only.
We begin by inspecting
\be
S_{4}\lsp A \rsp  \E  \frac{1}{4} g_2^{\, 2} \, \epsilon^{abc} \epsilon^{ade} \,
      \int \! d^4 x \,  A_{\mu}^{b}  \lrp x \rrp \, A_{\nu}^{c}  \lrp x \rrp \,
                        A_{\mu}^{d}  \lrp x \rrp \, A_{\nu}^{e}  \lrp x \rrp \, .
\label{S4}
\ee
The observation of this being a contraction of a Lorentz tensor
$\epsilon^{abc}A_{\mu}^{b}A_{\nu}^{c}$ with itself, could motivate a definition
of this tensor as a composite new field, and (since a glance at its gauge
behavior would quickly suggest its completion to the non-abelian
field-strength tensor $G^a_{\mu \, \nu}$) this would naturally lead to the
introduction of the $G^a_{\mu \, \nu}$ as new, {\em tensorial}\, field variables,
as envisaged in the field-strength formulation of Halpern\cite{Halp,Scha}.
But we might pair off the Lorentz indices the other way around and introduce
the Lorentz, or rather Euclidean, {\em scalar} composite fields
\be
\Xi^{\,a\,b} \lrp x \rrp   \Def   A_{\mu}^{a}  \lrp x \rrp  
                                       \, A_{\mu}^{b}  \lrp x \rrp \, ,
\label{Xidef}
\ee
which form a symmetric tensor in isospace, of mass dimension two, and
evidently positive semi-definite in the Euclidean. As a collection of the
scalar products between three vectors, this isotensor resembles what in
linear algebra is called a {\em Gramian matrix}, except that the number of
vectors does not fit the spatial dimension. At this point, in order to
avoid a rather repetitious later discussion, we anticipate that we will be
forced to consider the extension of (\ref{Xidef}) to a {\em four-dimensional}
matrix in extended isospace,
\be
X^{A\,B} \lrp x \rrp  \Def   A_{\mu}^{A}  \lrp x \rrp  
                                       \, A_{\mu}^{B}  \lrp x \rrp \, ,
\label{Xdef}
\ee
{\em in spite of}  the fact that only its $3 \times 3$
submatrix $\Xi$ appears in (the non-abelian part of) the action.
This is now a genuine Gramian matrix,
and it is well known\cite{Bern} that this matrix, at a certain Euclidean
point $x$, encodes the linear-dependence properties of the set of four
vectors at that point: it is nonsingular if and only if those vectors are
linearly independent. In that case, it is then also positive definite. We will
follow \cite{Halp} in assuming that this is in fact the generic case, i.\,e.\ that
a vanishing determinant of $X$ will at most occur along lower-dimensional
submanifolds of the $\mathrm{E}^4$ space that make zero contribution to the
action integral. For our purposes, $X$ is then a {\em generically positive
definite} $4 \times 4$ matrix field.                                           

The process we referred to as extracting a root now consists in invoking the
simple fact\cite{Bern} that a symmetric and positive semi-definite matrix
like $X$ can always be decomposed as
\be
X \lrp x \rrp  \E  Q \lrp x \rrp \, \cdot \, Q^T \lrp x \rrp \  ,
\label{Qdef}
\ee
with $T$ denoting a matrix transpose. If $X$ is nonsingular, so is $Q$, but
otherwise $Q$ is a general $4 \times 4$ matrix field, of mass dimension one
in $d = 4$, and with $16$ independent component fields. Note that this is now
a matrix factorization in the extended isospace that in contrast to eq.~(\ref{Xdef}) 
no more involves the Lorentz indices. Clearly the choice of $Q$
is not unique, as a given $Q$ can always be transformed, without changing $X$,
into $ Q \, \cdot \, O$ with $O$ an {\em orthogonal} $4 \times 4$ matrix. Thus
in order to parameterize $Q$, we may start with the simplest possibility,
$Q \E X^{\frac{1}{2}} $, where $X^{\frac{1}{2}}$ is, at each space point $x$,
the positive semi-definite matrix square root of $X$ and is, like $X$ itself,
a symmetric matrix. It thus provides ten independent component fields. Then
the general $Q$ can be represented as
\be
Q \lrp x \rrp  \E  X^{\frac{1}{2}} \lrp x \rrp \,
               \cdot \, \exp \lsp \, \Omega \lrp x \rrp \, \rsp \, ,
\label{Qpol}
\ee
where the orthogonal-matrix field is generated by
\be
\Omega^{A B} \lrp x \rrp  \E - \Omega^{B A}  \lrp x \rrp \, ,  
\label{Omega}
\ee
an {\em antisymmetric} and dimensionless $4 \times 4$ matrix field, which
provides the remaining six independent component fields. Thus $ \exp \lsp 
\Omega (x) \rsp $ is an element of SO(4), the real orthogonal group in four
dimensions, but it is to be noted that the rotation here takes place, not in
the Euclidean space, but rather in the four-dimensional extended isospace
referred to by the upper-case indices. Eq.~(\ref{Qpol}) represents what is
known as the right polar decomposition of $Q$, a decomposition of type modulus
$\times$ phase. The determination of $ \exp \lsp \Omega (x) \rsp $ by further
requirements will be the subject of the next section.                

%
Since $X$ has a distinguished upper-left $3 \times 3$ submatrix (\ref{Xidef})
that enters the \mbox{quadrilinear action} $S_4$, it seems natural to adopt
three-plus-one partitionings for both $X$ and its \mbox{``root'' $Q$, writing}
\bea            
\textnormal{\large $Q$}   \E  \lrp \begin{array}{cccc}
            \multicolumn{3}{c}{ \textnormal{\raisebox{-0.7mm}
                               {{\Large $\Psi$}  }} \quad }  & \chi   \\
            \multicolumn{4}{c}{ {} }                                  \\
            \multicolumn{3}{c}{ \eta^T \quad }         & \psi   \\ 
              \end{array} \rrp \  . 
\label{Qpart}
\eea        
Here $\chi^a \E Q^{a4} \ \left( a \E 1, 2, 3 \right) \,$ is a column $3$-vector,
$(\eta^T)^a \E Q^{4a} \ \left( a \E 1, 2, 3 \right) \,$ a collection of $3$
singlets, and $\psi \E Q^{44}$ a singlet in isospace,
but these designations are to be taken with
a grain of salt -- they do not, in these cases, imply homogeneous
transformation properties under local-gauge and BRS transformations. Here,
since all our scalars are defined through the gauge fields, they will inherit
in some form the {\em inhomogeneous} transformation laws of the latter. (We
collect the gauge/BRS variations in appendix A; they will typically
feature, in addition to an homogeneous piece, an inhomogeneous term involving
derivatives $\partial_{\mu} \theta^A (x)$ of the gauge functions $\theta^A$).

The $3 \times 3$ matrix $\Psi$ carries a reducible representation of isospin
that naively would be expected to reduce according to the dimension count
$ \mathbf{9} = \mathbf{1} \,+\, \mathbf{3} \,+\, \mathbf{5} $, but a glance
at eq.~(\ref{Psivar}) shows that in this case we rather have a set of three
(global) isotriplets:
\be
\mathbf{9} = \mathbf{3} \,+\, \mathbf{3} \,+\, \mathbf{3},
\label{Psired}
\ee
The $\chi$ entry provides a further (global) isotriplet. Similarly, the
last line of  (\ref{Qpart}) comprises four \mbox{(global)} isosinglets,
so the sixteen-dimensional representation $Q$ gets reduced according to
\be
 \mathbf{16} = 4 \, \times \, \mathbf{3} \,+\, 4 \, \times \, \mathbf{1} 
\label{Qred}
\ee
We reemphasize that because of the special gauge/BRS behavior of our
scalars, we may at most speak of {\em global} isomultiplets, i.\,e.\ multiplets
under spatially constant but not under local gauge transformations.

Using (\ref{Qpart}) and its transpose, the $X$ of eq.~(\ref{Qdef}) now gets
partitioned as 
\bea
X  \E  \lrp \begin{array}{cccc}
       \multicolumn{3}{c}{ \Psi \cdot \Psi^T + \chi \cdot \chi^T    }  &  
                        \  \Psi \cdot \eta   + \psi \cdot \chi           \\  
       \multicolumn{4}{c}{ {} }                                          \\
       \multicolumn{3}{c}{ (\Psi  \cdot \eta   + \psi \cdot \chi)^T }  &
                        \  \eta^T \cdot \eta + \psi^2
              \end{array} \rrp                                     
\label{Xpart}
\eea
(We employ the usual notation where $\Psi \cdot \Psi^T$ is a product
matrix while $\eta^T \cdot \eta$ is a scalar product of vectors) .
The point here is that the upper-left $3 \times 3 $ matrix $\Xi$ does not
appear simply as $ \Psi \cdot \Psi^T $, as it would if we had ``extracted a
root'' from $\Xi$ alone, but has an additional separable term $ \chi \chi^T $ 
that still preserves its (generically) positive definite character. In
terms of components, then,
\be
\Xi^{\,a\,b} \lrp x \rrp  \E  \Psi^{\,a\,c}(x) \, \Psi^{\,b\,c}(x) \, + \,
                            \chi^a(x) \, \chi^b(x) \, .
\label{Xicomp}
\ee
For later use, we note that a similar decomposition may also be given for
the complete four-dimensional matrix $X$: from eq.~(\ref{Xpart}) we have
\be
X  \E  P \cdot P^T \, + \, q \cdot q^T \  , 
\label{Xcomp}
\ee
where $P$ is the $4 \times 4$ matrix with three-plus-one partitioning
\bea            
\textnormal{\large $P$}   \E   \lrp \begin{array}{cccc}
            \multicolumn{3}{c}{ \textnormal{\raisebox{-0.7mm}
                               {{\Large $\Psi$}  }} \quad }  &  0     \\
            \multicolumn{4}{c}{ {} }                                  \\
            \multicolumn{3}{c}{ \eta^T \quad }         &  0     \\ 
                                     \end{array} \rrp \  , 
\label{Pmat}
\eea        
while $q$ denotes the $4$-isovector
\be
q  \E  \begin{pmatrix}
             \chi^1  \\
             \chi^2  \\
             \chi^3  \\
             \psi  
             \end{pmatrix}     \ .
\label{smallq}
\ee
The difference from eq.~(\ref{Xicomp}) is that obviously $\det P \, = \, 0$
and therefore
\be
\det ( P \cdot P^T ) \E 0 \  .  
\label{Pdet}
\ee
More precisely, $P$ and $P\,P^T$ are generically of rank three.

With these scalar-isomatrix fields, the quadrilinear part (\ref{S4}) of the
action can now be rewritten in either of two equivalent forms: the first,
\be
S_{4} \lsp Q,\,A \rsp  \E  \frac{1}{4} \, g_2^{\, 2} \, \int \! d^4 x \, 
A_{\mu}^{b}  \lrp x \rrp \, \left\{ \epsilon^{abc} \epsilon^{ade} \, 
\Xi^{\,c\,e} \lrp x \rrp \right\}  A_{\mu}^{d}  \lrp x \rrp \,  
\label{S4mix}
\ee
is, by eq.~(\ref{Xicomp}), bilinear in our scalar fields. It is thus a
scalar-gauge interaction of the ``seagull'' type. The second,
\be
S_{4} \lsp Q \rsp  \E  \frac{1}{4} \, g_2^{\, 2} \, \int \! d^4 x \,
\left\{ \epsilon^{abc} \epsilon^{ade} \, \Xi^{\,b\,d} \lrp x \rrp \, 
\Xi^{\,c\,e} \lrp x \rrp \,\right\}
\label{S4quart}
\ee
is a quadrilinear self-interaction of the scalar fields. Recall that these
two kinds of interactions appear {\em separately} in the standard-model Higgs
action, whereas here they are equivalent forms of the same term.

    The foregoing construction of the matrix field $Q$ makes essential use of
the theory operating in Euclidean space, since it is only there that the
scalar products in $X$ are positive definite. It may therefore not be
redundant to recall one of the basic tenets of Euclidean field theory:
analytic continuation to the Minkowskian domain is to be performed for the
final, c-number correlation functions (Schwinger functions) generated,
{\em not} in the equations of motion, or the path integrals, of the theory,
whose treatment must be performed entirely in the Euclidean. 
\vspace{0.5cm}

\section{The trilinear coupling}
\setcounter{equation}{0}
%

We have rewritten the four-fields term $S_4$ as an interaction between the
new scalars $Q$ and their ``parent'' gauge fields. We now wish to do the same
for the trilinear term $S_3$. This will turn out to also determine partially,
though not completely, the orthogonal-matrix factor $\exp(\Omega)$ of
eq.~(\ref{Qpol}). Of course, the gauge-invariant way of generating both types
of coupling is through (scalar contractions of) gauge-covariant derivatives,
but here we defer the introduction of a suitable concept of covariant
derivative to section 6, where it will be facilitated by the use of an
orthonormal vector system attached, in a sense, to the gauge fields. The
results of the present section will in fact lead directly to that very
convenient tool. 

      For physical reasons, we expect the trilinear term to assume the form 
of a coupling of the gauge field to scalar-field currents. We inspect the
$S_3$ term of (\ref{gaugact}), which again features only lower-case isospin
indices:
\be
S_3\lsp A \rsp  \E  \frac{1}{2} \, g_2 \, \int \! d^4 x \, 
   \lrp \partial_{\mu} A_{\nu}^{a} \, - \, \partial_{\nu} A_{\mu}^{a}  \rrp \,  
\lrp  \epsilon^{abc} \, A_{\mu}^{b}  \lrp x \rrp \, A_{\nu}^{c}  \lrp x \rrp \rrp . 
\label{S3}
\ee
With some index reshuffling, this can be rewritten as
\be
S_3\lsp A \rsp  \E  \frac{1}{2} \, g_2 \, \epsilon^{abc} \, \int \! d^4 x \,  
   \lsp  A_{\lambda}^b  \lrp x \rrp  \drl_{\!\!\mu} \, A_{\lambda}^c \lrp x \rrp
   \rsp \cdot \, A_{\mu}^{a} \lrp x \rrp  \, ,  
\label{S3drl}
\ee
where the usual notation, $A \drl B \E A(\partial B) - (\partial A)B$, has
been employed. The desired structure will emerge if we can choose our
matrix-of-scalars $Q$ such that
\bea
A_{\lambda}^B \lrp x \rrp  \drl_{\!\!\mu} \  A_{\lambda}^C \lrp x \rrp
        \E  \lsp Q  \drl_{\!\!\mu} \, Q^T \rsp^{\,B\,C} \  ,
\label{diffcond}
\eea
(of which, again, only the $3 \times 3$ submatrix $(B,C) \rightarrow (b,c)$
enters into (\ref{S3drl})), but at first sight this seems an unlikely
prospect: the freedom we have in $Q$ at this point consists in the six
independent components of $\Omega$, eq.~(\ref{Omega}), but while the two sides
of (\ref{diffcond}) are antisymmetric in their U(2) indices, the relation
is to hold for each of the four values of $\mu$ and thus implies $6 \times 4
\E 24$ conditions. Obviously, this can work only if substantial constraints
apply among the four $\mu$ values. It is all the more remarkable, and perhaps
a piece of insight into the structure of gauge theory, that such constraints
can be identified {\em and are in fact fulfilled}.

The key to bringing eq.~(\ref{diffcond}), which as a condition on $\Omega$ is
rather unwieldy, to a transparent form making the necessary constraints
evident, is the introduction of a quadruple of vector fields obtained by
{\em orthonormalizing the four gauge fields}. This is possible by means of the
Gram matrix $X$ of \mbox{(\ref{Xdef}):} $X$ being generically
nonsingular, so is its matrix root $ X^{\frac{1}{2}} $, and we can define
\be
N_{\mu}^A \lrp x \rrp  
 \Def  \, \lsp X(x)^{-\frac{1}{2}} \rsp^{\,A\,B} \, A_{\mu}^B  \lrp x \rrp
 \E   A_{\mu}^B  \lrp x \rrp \, \lsp X(x)^{-\frac{1}{2}} \rsp^{\,B\,A} \,  , 
\label{Nvect} 
\ee
the second equation holding due to the symmetry of $ X^{-\frac{1}{2}} $.
It is immediate to see that these form an orthonormal system:
\bea
N_{\mu}^A \lrp x \rrp  \, N_{\mu}^B \lrp x \rrp  & \E &
   \lsp X(x)^{-\frac{1}{2}} \rsp^{\,A\,C} \, \underbrace{ \lsp  
        A_{\mu}^C  \lrp x \rrp \, A_{\mu}^D  \lrp x \rrp \rsp }_{X^{CD}} 
   \lsp X(x)^{-\frac{1}{2}} \rsp^{\,D\,B}                   \nonumber    \\
{} & \E & \delta^{\,A\,B} \  . 
\label{Ortho} 
\eea
Moreover, in $4$-dimensional Euclidean space they are a complete set, which
translates as  
\be
N_{\mu}^A \lrp x \rrp \,  N_{\nu}^A \lrp x \rrp  \E  \delta_{\,\mu\,\nu} \  .  
\label{Comple} 
\ee
This orthonormalization process is known as L\"owdin orthogonalization\cite{Loew} 
and was first developed for applications in quantum chemistry.
In our context, the basis constructed is different at each point $x$ in
${\mathrm E}^4$. Thus we have a bundle of orthonormal frames, attached to all
points of space. The only change such a basis can undergo while maintaining
its orthonormality is a rigid rotation, so $N$ bases at different points
of ${\mathrm E}^4$ differ at most by an SO(4) rotation. Definition
(\ref{Nvect}) may be turned around to give an expansion of the $A(x)$ field
in terms of the orthonormal system at $x$: 
\be
A_{\mu}^B \lrp x \rrp \  \E  \lsp X(x)^{\frac{1}{2}} \rsp^{\,B\,C}
\,                          N_{\mu}^C \lrp x \rrp \,  .  
\label{Arep} 
\ee
Using this expansion on the l.h.s.\ of eq.~(\ref{diffcond}), we have
\bea
A_{\lambda}^B \lrp x \rrp  \drl_{\!\!\!\mu} \  A_{\lambda}^C \lrp x \rrp
& \E & \lsp X^{\frac{1}{2}} \rsp^{\,B\,D}  \, \lsp N_{\lambda}^D  \drl_{\!\!\!\mu}
 N_{\lambda}^E \rsp \,  \lsp X^{\frac{1}{2}} \rsp^{\,E\,C}                   \non
& + & \left\{ N_{\lambda}^D \, N_{\lambda}^E \right\} \, \lsp \,\lrp X^{\frac{1}{2}} 
       \rrp^{\,B\,D} \drl_{\!\!\!\mu}  \lrp X^{\frac{1}{2}} \rrp^{\,E\,C} \, \rsp \  .
\label{ldiffcond}
\eea
On the other hand the r.h.s.\ of eq.~(\ref{diffcond}), upon using the
polar decomposition (\ref{Qpol}) of $Q$, becomes
\bea
\lsp Q  \drl_{\!\!\!\mu} \, Q^T \rsp^{\,B\,C} 
& \E &  \lsp X^{\frac{1}{2}} \rsp^{\,B\,D} \, \lsp {\mathrm e}^{\Omega} 
        \drl_{\!\!\!\mu} {\mathrm e}^{-\Omega} \rsp^{\,D\,E} \,
        \lsp X^{\frac{1}{2}} \rsp^{\,E\,C}                        \non
&  + &  \lsp X^{\frac{1}{2}} \drl_{\!\!\!\mu}  X^{\frac{1}{2}} \rsp^{\,B\,C} \  .
\label{rdiffcond}
\eea
Orthonormality turns the curly bracket in eq.~(\ref{ldiffcond}) into a
$\delta^{\,D\,E}$, so the second lines of both sides drop out from the
condition. Note that this simplification, which is crucial to the entire
subsequent development, would not happen if we were
to scale both sides of eq.~(\ref{diffcond}) by different numerical factors.
The result, as desired, is purely a condition on ${\mathrm e}^{\Omega}$:
\be
\lsp {\mathrm e}^{\Omega} \drl_{\!\!\!\mu} {\mathrm e}^{-\Omega} \rsp^{\,B\,C}
\E   N_{\lambda}^B \drl_{\!\!\!\mu} N_{\lambda}^C \  .
\label{econd}
\ee
From the trivial relations
\be
\partial_{\mu} \lrp {\mathrm e}^{\Omega} \, {\mathrm e}^{-\Omega} \rrp \E 0 \  , 
\qquad \qquad \partial_{\mu} \lrp N_{\mu}^A \, N_{\mu}^B \rrp \E 0 \  ,
\label{triv}
\ee
it follows that on both sides of (\ref{econd}), the second (minus) term of
the $\drl_{\!\!\!\mu}$ construct equals the first -- in other words, both sides
are antisymmetric matrices in the U(2) space, and are thus structurally
compatible. We may then simplify condition (\ref{econd}) to read
\be
\partial_{\mu} \, {\mathrm e}^{\Omega} \E\ {\mathcal M}_{\mu} \  
                 {\mathrm e}^{\Omega} \ ,
\label{diffeq}
\ee
with the $4 \times 4$ matrices ${\mathcal M}_{\mu}$ given by
\bea
\left( {\mathcal M}_{\mu} \right)^{\,A\,B} \lrp x \rrp  & \E & \frac{1}{2} \, 
     \lsp -N_{\lambda}^A \lrp x \rrp \drl_{\!\!\!\mu} N_{\lambda}^B \lrp x \rrp 
     \rsp                                                               \\
   & \E  &  -N_{\lambda}^A \lrp \partial_{\mu} N_{\lambda}^B \rrp
     \E   \ \lrp \partial_{\mu} N_{\lambda}^A \rrp N_{\lambda}^B \  .
\label{Mmu}
\eea
The latter two forms are again valid because of (\ref{triv}).

It is now evident which constraints must be met for eq.~(\ref{diffcond})
to be solvable by a suitable choice of $\Omega$ in (\ref{Qpol}): these
constraints are identical with the {\em compatibility conditions} for
the system of four partial differential equations (\ref{diffeq}):
\be
  \partial_{\nu} \lrp {\mathcal M}_{\mu} \,{\mathrm e}^{\Omega} \rrp  \E
  \partial_{\mu} \lrp {\mathcal M}_{\nu} \,{\mathrm e}^{\Omega} \rrp  \  .
\label{compa}
\ee
If these are fulfilled, the four right-hand sides of (\ref{diffeq}) are in
fact first partial derivatives of a single antisymmetric $4 \times 4$ matrix
quantity, with six independent components, so the freedom in $\Omega$ is just
sufficient for building that quantity.  Upon working out these conditions\cite{Zwill}, 
one finds
\be
\partial_{\mu} \, {\mathcal M}_{\nu} \, - \, \partial_{\nu} \, {\mathcal M}_{\mu}
 \, - \lsp {\mathcal M}_{\mu}\, , \, {\mathcal M}_{\nu} \rsp  \E  0  \  ,
\label{Mmucond}
\ee
the brackets denoting a matrix commutator. This means that 
$-{\mathcal M}_{\mu}$, when viewed as a matrix-valued gauge potential,
produces zero field strength. (It must then be a ``pure gauge'', which is
exactly what eq.~(\ref{diffeq}) is saying). Now compute
\be
\lrp \partial_{\mu} \,{\mathcal M}_{\nu}\, - \, \partial_{\nu}\,{\mathcal M}_{\mu}
\rrp^{\,A\,B}  \E \,-\, \lrp \partial_{\mu} \, N_{\lambda}^A \rrp \,
                      \lrp \partial_{\nu} \, N_{\lambda}^B \rrp 
                  +\, \lrp \partial_{\nu} \, N_{\lambda}^A \rrp \,
                      \lrp \partial_{\mu} \, N_{\lambda}^B \rrp \  , \qquad 
\label{rotmu}
\ee
\be
-\, \lsp {\mathcal M}_{\mu} , \,{\mathcal M}_{\nu} \rsp^{\,A\,B}   
\E  \lcp N_{\kappa}^C \, N_{\lambda}^C \rcp    \cdot
                \lsp \lrp \partial_{\mu} \, N_{\kappa }^A \rrp \,
                     \lrp \partial_{\nu} \, N_{\lambda}^B \rrp 
                 -\, \lrp \partial_{\nu} \, N_{\lambda}^A \rrp \,
                     \lrp \partial_{\mu} \, N_{\kappa }^B \rrp \rsp  \   .
\label{commu}
\ee
This time it is {\em completeness} of the $N$'s, eq.~(\ref{Comple}), that
turns the curly bracket in (\ref{commu}) into $ \delta_{\kappa \, \lambda} $
and, remarkably, makes the two quantities cancel. It now becomes obvious
why we had to work with the four-gauge-vectors formulation, rather than
with the non-abelian fields alone: had we proceeded from the $3$-dimensional
$\Xi$ of eq.~(\ref{Xidef}), we would have obtained, from the analog of eq.
(\ref{Nvect}), a set of {\em three} orthonormal vectors that would have been
incomplete in four Euclidean dimensions. It is interesting that in the present
context, the additional U(1) is not simply an empirical fact but is forced
upon us by the formalism.

An explicit form of the solution of system (\ref{diffeq}) will not be
absolutely necessary in the following, but is convenient to have. Since
$ {\mathcal M}_{\mu} $ is both $x$-dependent and a noncommuting quantity, such
a solution can in any case be exhibited only in a formal sense, as a
path-ordered exponential:
\be
{\mathrm e}^{\Omega}[A, \, {\mathcal C}; \, x)   \E    O[A, \, {\mathcal C}; 
                                      \, x) \   {\mathrm e}^{\Omega_0[A]} \  ,
\label{expinit}                                                        
\ee
where
\be
O[A, \, {\mathcal C}; \, x)  \E   \mathcal{P} \, \lcp
         \exp \, \lsp - \frac{1}{2} \, \int_{\mathcal{C}(x_0,x)} ds_{\mu}
         \lrp N_{\lambda} \lrp s \rrp \drl_{\!\!\!\mu} \, N_{\lambda} \lrp s \rrp 
         \rrp     \rsp      \rcp \  ,
\label{pathord}
\ee
so that $O[A, \, {\mathcal C}; \, x_0)  \E  \Id_4$. Here we employ
the obvious matrix notation $N_{\lambda} \drl_{\!\!\!\mu} \  N_{\lambda}$, where
\be
( \, N_{\lambda}   \drl_{\!\!\!\mu} \, N_{\lambda} \, )^{\,A\,B}  \E
     N_{\lambda}^A \drl_{\!\!\!\mu} \, N_{\lambda}^B  \,   .     
\label{NNmat}
\ee
Eqs.~(\ref{expinit}/\ref{pathord}) do, however, illustrate a number of
qualitative features. First, note the presence of a matrix-valued 
``integration constant'', $ {\mathrm e}^{\Omega_0} $, with $\Omega_{0}[A] = 
\Omega[A;x=x_0)$, a spatially constant SO(4) isorotation, which by
convention we have written as a right-hand factor. Thus, we still have not
pinned down the matrix field $Q$ uniquely, but are left with a {\em residual
freedom} encoded in one constant orthogonal matrix
            
Second, since the integrand $ds_{\mu} \cdot \mathcal{M}_{\mu}$
is a Lorentz (or Euclidean) scalar, $Q$ continues to be defined in terms
of scalars alone.

Third, upon expanding the path-ordered exponential one obtains a series
of multiple integrals with increasing numbers of the bilinear (\ref{NNmat}),
which may be viewed as a kind of coherent-state expansion, as mentioned
in section 1.

Fourth, since $O[A; \, {\mathcal C}; \,x)$ involves a path ${\mathcal C}$,
which we always take from a conventional starting point $x_0$ to $x$, it is
{\em prima vista} a nonlocal functional of the original gauge fields $A$,
in apparent contrast to the $X^{\frac{1}{2}}$ factor in (\ref{Qdef}), which
is built purely from the fields at its own argument $x$. Since the
${\mathcal M}_{\mu}$ in the integral is not a pure gradient, this path
dependence seems to be unavoidable. The \mbox{(provisional)} functional
notation in eq.~(\ref{expinit}) refers to this feature.

However, such a conclusion would not do justice to the nonabelian nature of
the expression, as encoded in the path ordering $\mathcal{P}$. To examine
this question, we first parameterize $\mathcal C$ as
\be
\mathcal{C}(0,x)  \E  \lcp s_{\mu}(x,t) \, : \, \mu \E 1,2,3,4; \quad t=0 \,
\ldots \, 1, \quad s(x,0) \E x_0, \, s(x,1) \E x \rcp  \  . 
\label{spath}
\ee
The path ordering $\mathcal{P}$ now turns into a time ordering $\mathcal{T}$
with respect to the path parameter $t$. The tool of choice is then a quantity
that interpolates $O[A; \, {\mathcal C}; \, x)$ in this curve parameter\cite{BDJ} ,
\be
U \lsp {\mathcal C}; \, x, \, t \leftarrow 0 \rrp  \E  \mathcal{T}
 \, \lcp \exp \, \lsp \int_{0}^{t} \, dt^{\prime} \, \frac{\partial \, s_{\mu}
(x,t^{\prime}) }{\partial \, t^{\prime}} \, {\mathcal M}_{\mu} \lrp s(x,
t^{\prime}) \rrp  \rsp  \rcp .
\label{Umatfunc}
\ee
$U$ is a unitary-matrix function obeying the differential equation
\be
\frac{\partial}{\partial \, t} \, U \lsp {\mathcal C}; \, x, \, t \leftarrow 0
\rrp  \E  \frac{\partial \, s_{\mu} (x,t) }{\partial \, t} \, {\mathcal
 M}_{\mu} \lrp s(x,t) \rrp \,   U \lsp {\mathcal C}; \, x, \, t \leftarrow 0 
\rrp  ,
\label{Udgl}
\ee
and the boundary conditions,
\be
U \lsp {\mathcal C}; \, x, \, 0 \leftarrow 0 \rrp \E \Id_4  ; \qquad 
U \lsp {\mathcal C}; \, x, \, 1 \leftarrow 0 \rrp \E O[\,{\mathcal C}; \, x)
\label{Ubound}
\ee
where in the last term we omitted the functional dependence on $A$.
(Notation: $\Id_n$ and $\mathrm{tr}_n$ are the unit matrix and trace,
respectively, in $n$ isodimensions, for $n = 2, 3, 4$). The infinitesimal
change in $U$,
\be
\delta_{{\mathcal C}} U \Def
U \lsp {\mathcal C}+\delta {\mathcal C}; \,x,\,t \leftarrow 0 \rrp \, - \,
U \lsp {\mathcal C}; \,x,\, t \leftarrow 0 \rrp \  ,
\label{Uvar}
\ee
produced by an infinitesimal change in the path,
\be 
{\mathcal C} \, \rightarrow \, {\mathcal C}\,+\,\delta {\mathcal C} \, :
\qquad  s_{\mu}(x,t) \, \rightarrow \,  s_{\mu}\,+\,\delta s_{\mu}(x,t);
\qquad \delta s_{\mu}(x,0) \E \delta s_{\mu}(x,1) \E 0 \  ,    
\label{pathvar}
\ee
is given by a standard formula for ${\mathcal T}$-ordered functionals\cite{Quot} ,
\bea
\delta_{{\mathcal C}} \, U  &  \E  &  \int_{0}^{1} \,  U \lsp s; \,x,\, t 
\leftarrow \tau \rrp \, \lcp \frac{\delta}{\delta s_{\nu}(\tau)} \lsp
\frac{\partial s_{\mu}}{\partial \tau} \, {\mathcal M}_{\mu} \lrp s(\tau) \rrp
\rsp \rcp  U^{-1} \lsp s; \,x,\, t \leftarrow \tau \rrp \, \delta
 s_{\nu}(\tau) \,  d \tau                                  \non
\qquad  &  \cdot &  U \lsp s; \,x,\,t \leftarrow 0 \rrp \  ,
\label{Tordvar}
\eea
where
\be
U^{-1} \lsp s; \,x,\, t \leftarrow \tau \rrp  \E
U \lsp s; \,x,\, \tau \leftarrow t \rrp   \  .
\label{Uinv}
\ee
By spelling out the bracketed functional derivative, applying partial
integration, using the differential equation (\ref{Udgl}) as adapted to the
variable $\tau$, and setting $t \E 1$, one finally obtains a result derived
via a different route in \cite{Corr}:
\bea
\lcp \delta_{{\mathcal C}} O \lsp {\mathcal C}; \, x \rrp \rcp \, O^T \lsp
{\mathcal C}; \, x \rrp & \E & - \, \int_{0}^{1} \, d \tau \, U \lsp s; \,x,
\, 1 \leftarrow \tau \rrp \, \lcp \frac{\partial}{\partial s_{\mu}} \, 
{\mathcal M}_{\nu} \, - \, \frac{\partial}{\partial s_{\nu}} \, 
{\mathcal M}_{\mu} \, - \lsp {\mathcal M}_{\mu}\, , \, {\mathcal M}_{\nu} \rsp
\rcp_{s(\tau)}                                                 \non
& \times & U^{-1} \lsp s; \,x,\, 1 \leftarrow \tau \rrp \, \cdot
\frac{\partial s_{\mu}(x,\tau)}{\partial \tau} \, \delta s_{\nu}(x,\tau) \  .
\label{Opathvar}
\eea
A glance at eq.~(\ref{Mmucond}) then shows that $\delta_{{\mathcal C}} O \E 0$.
Contrary to appearances, the matrix $O$ is therefore not a path functional;
it continues to depend functionally on the gauge fields, which we may call
its {\em parents}, but otherwise is just a local function of $x$.

         In the following, when we wish to refer to a completely specified
quantity, we will therefore choose, without loss of generality, the straight
path $s_{\mu}(x,t) \E  t \cdot ( x \, - \, x_0 )_{\mu}$ , in which case
\be
O[A;x) \E  \mathcal{T} \, \lcp
       \exp \, \lsp - \frac{1}{2} \, ( x \, - \, x_0 )_{\mu} \, 
       \int_{0}^{1} \, dt \,  \big( N_{\lambda} \lrp s \rrp \drl_{\!\!\!\mu} \,
        N_{\lambda} \lrp s \rrp \big)_{s=t(x-x_0)}   \rsp   \rcp \   .
\label{straight}
\ee
This orthogonal-matrix function obeys the initial condition $O[A;x=x_0)
\E \Id_4$. 

     With the possibility of eq.~(\ref{diffcond}) now assured, we may
finally rewrite (\ref{S3drl}) in the form envisaged, that of a set of
scalar-field currents coupling to the SU(2) gauge fields: 
\be
S_3 \lsp A \rsp  \E  \frac{1}{2} \,  g_2 \, \epsilon^{abc} \, \int \! d^4 x \    
[ Q  \drl_{\!\!\mu} \, Q^T ]^{\,b\,c} \cdot \, A_{\mu}^{a} \lrp x \rrp \,  .  
\label{S3QQT}
\ee
In terms of the component fields of eq.~(\ref{Qpart}), the $Q$ current gets
partitioned as 
\bea
 Q  \drl_{\!\!\mu} \, Q^T   \E   \left( \begin{array}{cccc}
       \multicolumn{3}{c}{ \Psi \drl_{\!\!\!\mu} \, \Psi^T
                         + \chi \drl_{\!\!\!\mu} \, \chi^T }  &  
   \                       \Psi \drl_{\!\!\!\mu} \, \eta  
                         + \psi \drl_{\!\!\!\mu} \, \chi                   \\  
       \multicolumn{4}{c}{ {} }                                          \\
       \multicolumn{3}{c}{(\Psi \drl_{\!\!\!\mu} \, \eta
                         + \psi \drl_{\!\!\!\mu} \,  \chi)^T }  &  \  0
                                      \end{array} \right)             
\label{Currpart}
\eea
The last entry obtains because $ \eta^a \drl_{\!\!\!\mu} \,\eta^a  \E  \psi
\drl_{\!\!\!\mu} \,\psi \E 0 $. One sees that the upper-left submatrix that
enters into (\ref{S3QQT}),
\be
[ Q(x) \drl_{\!\!\mu} \, Q^T(x) ]^{\,b\,c}  \E  
[ \, \Psi(x)   \drl_{\!\!\!\mu} \, \Psi^T(x) \, ]^{\,b\,c} \, + \,
     \chi^b(x) \drl_{\!\!\!\mu} \, \chi^c(x) \   , 
\label{QdQT3}
\ee
again involves not only the $3 \times 3$ matrix $\Psi$ but has an additional,
separable $\chi$ term.
\vspace{0.4cm}

\section{Constant orthonormal frame}
\setcounter{equation}{0}
%
This section is a technical interlude which does, however, lead up to
something conceptually \mbox{significant:} the interpretation, in
eq.~(\ref{QprimAn}) below, 
of the scalars $Q^{A \, B}$ as {\em projections} of the gauge-field
vectors onto a constant orthonormal basis that is defined by the fields
themselves.

Eq.~(\ref{diffcond}) has further consequences that are best formulated
by introducing the modified L\"owdin basis vectors
\be
\tilde{N}_{\lambda}^A  \Def  \lrp {\mathrm e}^{-\Omega} \rrp^{\,A\,B} N_{\lambda}^B
                     \E    N_{\lambda}^B \lrp {\mathrm e}^{\Omega} \rrp^{\,B\,A}. 
\label{Ntilde}
\ee
Since these are obtained from the $N's$ by an orthogonal transformation in
extended isospace, they too form an orthonormal and complete system:
\bea
\tilde{N}_{\mu}^A  \, \tilde{N}_{\mu}^B   & \E &  \delta^{\,A\,B}    \,  ,
                  \label{tildo}                                        \\
\tilde{N}_{\mu}^C  \, \tilde{N}_{\nu}^C   & \E &  \delta_{\,\mu\,\nu} \,  . 
                  \label{tildc}
\eea
At this point they seem to depend on  $x$, like the $N$ vectors. 
By eqs.~(\ref{Nvect}) and (\ref{Qpol}), they can be expressed as
\be
\tilde{N}_{\mu}^A   \E
 \lrp Q(x)^{-1} \rrp^{\,A\,B} \, A_{\mu}^B (x)  \E  
 A_{\mu}^B (x) \lrp Q^T(x)^{-1} \rrp^{\,B\,A} \,  . 
\label{NQA} 
\ee
Using orthonormality, this can be turned around to give an instructive
representation of the $Q$-matrix elements:
\be
Q^{\,A\,B}  \E  A_{\lambda}^A \, \tilde{N}_{\lambda}^B \ .  
\label{QAN} 
\ee
While once again presenting them as Lorentz scalars, this formula identifies
the $Q^{\,A\,B}$ as the sixteen {\em projections} of the four gauge fields
onto the four orthonormal $\tilde{N}$ vectors.

The salient property of the latter emerges when relation (\ref{diffcond}),
now established, is combined with the derivative of (\ref{Qdef}),
\be
 A_{\lambda}^B  \lrp \partial_{\mu} A_{\lambda}^C  \rrp \,+\, \lrp \partial_{\mu}
 A_{\lambda}^B  \rrp A_{\lambda}^C      
 \E  \lsp Q \lrp \partial_{\mu} Q^T \rrp \,+\, \lrp \partial_{\mu} Q \rrp Q^T
 \rsp^{\,B\,C}.
\label{sumcond}
\ee
By adding and subtracting these one finds the relation
\be
\lsp \lrp \partial_{\mu} Q \rrp Q^T \rsp^{\,B\,C}  \E 
\lrp \partial_{\mu} A_{\lambda}^B \rrp A_{\lambda}^C  \   ,
\label{singcond} 
\ee
and its transpose. Or, by combining this with eq.~(\ref{NQA}),
\be
\lrp \partial_{\mu} Q \rrp^{\,B\,C}  \E 
\lrp \partial_{\mu} A_{\lambda}^B \rrp \tilde{N}_{\lambda}^C  \  .
\label{dQAN}
\ee
Upon comparing this with the partial derivatives of eq.~(\ref{QAN}), we
conclude that
\be
A_{\lambda}^B \lrp \partial_{\mu} \tilde{N}_{\lambda}^C \rrp  \E  0  \  
\qquad ( \  \mathrm{all} \ \mu, \ B, \ C, \  )
\label{dtildN}
\ee
Thus each partial derivative of an $\tilde{N}$ vector is orthogonal to
all gauge vectors. The latter being generically linearly independent, we
conclude that, generically, the $\tilde{N}$'s are {\em spatially constant}:
\be
\partial_{\mu} \, \tilde{N}_{\lambda}^A   \E  0  \qquad \lrp \mathrm{all} \  
\mu, \, \lambda, \, A \rrp \  .
\label{tNconst}
\ee
This somewhat unexpected property shows that the transformation of eq.
(\ref{Ntilde}) has undone the $x$-dependent $4$-rotation by which the
original L\"owdin frames at various points differed. It is clear that the
$x$-independence property simplifies calculations in these frames 
significantly.

Of course, we cannot exclude that the $\tilde{N}$'s, while spatially
constant in a given gauge-field configuration
\be
\mathcal{A}  \Def  \lcp A_{\mu}^C (x) \  | \  \mathrm{all\ x\ in\ } E^4 ;
                \,  \mu, \, C \E 1....4 \rcp \  ,  
\label{config}
\ee
may still be different for different field configurations. But such a
variation is strongly restricted by condition (\ref{tNconst}) and
orthonormality. Since $\tilde{N}$'s must be constant separately in each of
two different
configurations $\mathcal{A}, \, \mathcal{A}'$, they can differ between these
only by a constant rotation, which because of the perfect symmetry between 
eqs.~(\ref{tildo}) and (\ref{tildc}) may be taken to be an SO(4) element on
either Euclidean or extended isospace. After selecting some reference
configuration $\mathcal{A}_0$, we then have for any $\mathcal{A}$,
\be
\tilde{N}_{\lambda}^B \lsp \mathcal{A} \rsp  \E   
          \lrp {\mathrm e}^{-\Gamma_{0}[\mathcal{A}]} \rrp^{\,B\,C}  \,
\tilde{N}_{\lambda}^C \lsp \mathcal{A}_0 \rsp  \  ,
\label{tildiff}
\ee
with $-\Gamma_{0}[\mathcal{A}]$ a constant, antisymmetric generator matrix
typical of configuration $\mathcal{A}$. Now recall that $\tilde{N}$, through
the ${\mathrm e}^{-\Omega}$ factor in its definition (\ref{Ntilde}), contains
just a matching freedom in the form of the integration-constant matrix in
eq.~(\ref{expinit}). Writing (\ref{tildiff}) with (\ref{Ntilde}) as
\be
{\mathrm e}^{\Gamma_{0}[\mathcal{A}]}  \tilde{N}_{\lambda} \lsp \mathcal{A} \rsp
\E  \lrp {\mathrm e}^{\Gamma_{0}[\mathcal{A}]} \, {\mathrm e}^{-\Omega_0[\mathcal{A}]}
\rrp \, \lrp O^{T}[\mathcal{A};x) \, N_{\lambda}[\mathcal{A};x) \rrp
\E  \tilde{N}_{\lambda} \lsp \mathcal{A}_0 \rsp  \  ,
\label{peeloff}
\ee
and choosing the ``integration constant'' as
\be
\Omega_{0}[\mathcal{A}]  \E  \Gamma_{0}[\mathcal{A}]  \  ,
\label{Omnull}
\ee
we see that the once-more-modified L\"owdin orthonormal system,
\be
n_{\mu}^B  \Def  \lrp O^{T}[A;x) \rrp^{\,B\,C} \, N_{\mu}^C[A;x)
          \E  \lrp {\mathrm e}^{\Gamma_{0}} \rrp^{\,B\,D} \,\tilde{N}_{\mu}^D \,,
\label{nvect}
\ee
is always equal to $ \tilde{N}_{\mu}^B \lsp \mathcal{A}_0 \rsp $ for the
reference configuration, and contrary to appearances is therefore independent
of both $x$ {\em and of the gauge-field configuration} $\mathcal{A}$:
\be
\partial_{\lambda} \, n_{\mu}^A \E 0 \, ;  \qquad \quad  \delta \, n_{\mu}^A 
\E 0 \, \quad \mathrm{under}  \quad  \mathcal{A} \,
       \longrightarrow \, \mathcal{A} \,+\, \delta \mathcal{A} \  .
\label{ninv}
\ee
Since the latter change may arise from a gauge or BRS transformation,
the $n$ system is, in particular, a {\em gauge and BRS invariant}.

     A slight modification of the $Q$ matrix,
\be
Q'(x)  \E  Q(x) \, {\mathrm e}^{-\Gamma_0} \, \E  X^{\frac{1}{2}}(x) \, O(x) \, ,
\label{Qprime}
\ee
which is still of the general form of eq.~(\ref{Qpol}), then has matrix
elements
\be
( Q' )^{\,A\,B}  \E  A_{\lambda}^A \, n_{\lambda}^B \  ,       
\label{QprimAn} 
\ee
which are the scalar projections of the $A$ vectors onto the $n$ vectors -- 
a pleasantly simple, geometric interpretation.

We emphasize that for our task of reformulating the gauge action, it is not
necessary to actually construct quantities such as $\Gamma_{0}[\mathcal{A}]$
or $ \tilde{N}_{\lambda} \lsp \mathcal{A}_0 \rsp $ explicitly; it suffices to
know that a BRS-invariant basis $n_{\lambda}^B$ exists and that the $Q'$
matrix obeying eqs.~(\ref{Qprime}) and (\ref{QprimAn}) continues to fulfill
relations (\ref{Qdef}) and (\ref{diffcond}). In the main text we therefore 
drop the prime, referring to it again as $Q$.

     Note that there is a difference, both conceptual and practical, between
these projections and the components of $A$ in the {\em arbitrary and fixed}
coordinate system tacitly understood when writing them as $A_{\mu}^C$. These
are not scalars but vector components. The orthonormal $n$-frame
vectors, while spatially constant and independent of field configuration,
do change their {\em components} in the arbitrary external basis in the same
way as the $A$ vectors do when that basis is changed, in such a way as to
keep the contraction over Lorentz indices in  eq.~(\ref{QprimAn}) unchanged.
In this sense the $Q$ elements are indeed scalar fields.
\vspace{0.4cm}

\section{Action in terms of Q's}
\setcounter{equation}{0}
%
We seek to rewrite the generating functional of Euclidean correlation
functions,
\bea
G \lsp J, \, K \rsp   & \E &  \frac{ Z \lsp J, \, K \rsp }
                                   { Z \lsp 0, \, 0 \rsp }
\label{Gfrac}                                                         \\
Z \lsp J, \, K \rsp   & \E &  \int \! \mathcal{D}[A] \mathcal{D}[c,\bar c]
          \  \mathrm{e}^{-S_E[A,\bar c, c] \, + \, (J,A) \, + \, (K,\,Q[A]) }
\label{pathint}                                                        
\eea
in a way that facilitates a study of the dynamics of either the scalar $Q$
fields among themselves, or of their coupling to their parent gauge fields.
In eq.~(\ref{pathint}), since  we are not interested here in ghost amplitudes,
we have omitted sources for the ghost fields, but we do carry sources
$K \E \{\,K_{\,A\,B}\,\}$ for easier generation of correlations of the $Q$
scalars, which at this point are viewed as functionals $Q[A]$ of the $A$'s.
Like the gauge-field sources $J$, they are of mass dimension three.
Standard scalar-product notation such as $(J,A) \E \int d^4x J_{\mu}^C (x)
A_{\mu}^C (x)$ has been employed. Here and in section 8, we
examine two ways of reformulating $Z$ using the scalar fields. 

First, we might consider a {\em complete reformulation in terms of scalars},
which formally turns gauge theory into a theory of scalar fields interacting
through tri- and quadrilinear couplings. For this, one uses the inversion of
eq.~(\ref{QprimAn}) via the completeness relation for the $n$ basis,
\be
 A_{\mu}^B   \E   Q^{\,B\,C} \, n_{\mu}^C \  . 
\label{AQn} 
\ee
and in particular, for $B = b = 1,\,2,\,3$ ,
\be
 A_{\mu}^b   \E   \Psi^{\,b\,c} \, n_{\mu}^c \  + \  \chi^{\,b} \, n_{\mu}^4 \  . 
\label{APsichin} 
\ee
Thanks to the configuration independence of the $n$ basis, this relation is
effectively linear, and the functional Jacobian is a constant that drops out
of the ratio (\ref{Gfrac}). The $J$-source term may now be omitted from eq.
(\ref{pathint}), and the $[A]$ argument from the $Q$-source term. The rewriting
uses, in addition to eqs.~(\ref{S4quart}) and (\ref{diffcond}), the two pieces
of the integrand in the bilinear term (\ref{S2}),
\bea
\frac{1}{2} \, \lrp  A_{\mu}^C \, , \delta_{\,\mu \,\nu}\,(-\partial_{\lambda}
\partial_{\lambda} )\, A_{\nu}^C \rrp
\, & = & \,\frac{1}{2} \,\mathrm{tr}_4 \, \lrp \partial_{\lambda} Q \, , \, 
       \partial_{\lambda} Q^T  \rrp \  ,   
\label{S2diag}                                                           \\     
\frac{1}{2} \, \lrp  A_{\mu}^C \, , (\partial_{\mu} \,\partial_{\nu} )\, 
A_{\nu}^C  \rrp \E  
& - & \frac{1}{2} \, \mathrm{tr}_4 \, \lrp \partial_{\mu} Q \, , \, (n_{\mu} 
\cdot n_{\nu}^T) \ \partial_{\nu} Q^T \rrp \  .
\label{S2long}
\eea
In the latter equation, we note the occurrence of the invariant,
orthonormal-basis vectors $n_{\mu}^C$, introduced in eq.~(\ref{nvect}),
in the tensorial combination
\be
\lrp n_{\mu} \cdot n_{\nu}^T \rrp^{\,B\,C}  \Def  n_{\mu}^B \, n_{\nu}^C \  .
\label{ncrossn}
\ee
Eq.~(\ref{S2diag}) shows that the ``diagonal'' piece alone of $S_2$ already
yields the complete, normalized kinetic term for the scalars:
\be
\frac{1}{2} \,\mathrm{tr}_4 \, \lrp \partial_{\lambda} Q \, , \, \partial_{\lambda}
Q^T \rrp  \E  \frac{1}{2} \,\mathrm{tr}_3 \, \lrp \partial_{\lambda} \Psi \, ,
\, \partial_{\lambda} \Psi^T \rrp \, + \,  \frac{1}{2} \, \lsp \lrp 
\partial_{\lambda} \chi^a \, ,\, \partial_{\lambda} \chi^a \rrp \, + \, 
\lrp \partial_{\lambda} \eta^b \, ,\, \partial_{\lambda} \eta^b \rrp \rsp \ + \  
\frac{1}{2}\, \lrp \partial_{\lambda} \psi \, ,\, \partial_{\lambda} \psi \rrp. 
\label{kinet} 
\ee
On the other hand, eq.~(\ref{S2long}) for the ``longitudinal'' piece of $S_2$
illustrates the fact that in this formulation, uncontracted $n$-basis vectors
remain in some, though not all, terms of the action. This feature will reoccur,
in particular, in the trilinear terms such as $S_3$ and the fermion-gauge and
ghost-gauge interactions. For these, the formulation is intrinsically
awkward, since it tends to obscure the vector nature of these couplings. 
The explicit $n$ vectors present an artificial element, conceptually different
from but practically somewhat similar to the constant external vectors that
one introduces in the axial gauges for vector gauge theories. The piece
(\ref{S2long}) with its $n$-tensor (\ref{ncrossn}) is best treated together
with the gauge-fixing term in $S_{GFG}$.

By contrast, the ``all-$Q$'s'' formulation is quite suitable for an inspection
of the $S_4$ term, in the form of eq.~(\ref{S4quart}), in which all $n$'s 
drop out by contraction. Upon applying, as usual, the SU(2)-specific relation
\be
\epsilon^{\,a\,b\,c} \epsilon^{\,a\,d\,e} \E   \delta^{d\, b} \, \delta^{e\, c} 
                         \  - \  \delta^{d\, c} \, \delta^{e\, b}  \  , 
\label{epseps} 
\ee
that term assumes the form,
\be
S_{4}  \E  \frac{1}{4} \, g_2^{\, 2} \, \int \! d^4 x \, \lcp   \lsp 
\mathrm{tr_3} \, \Xi(x) \rsp^2  \  - \  \mathrm{tr_3} \, \lsp \Xi(x)^2 \rsp
\rcp  \  . 
\label{S4tr} 
\ee
Note that the coupling constant here is the usual $g_2$ of the non-abelian
sector; in the present context there is {\em no need to introduce an extra
constant (usually denoted $\lambda$) for the quartic scalar term}. For a
positive-definite matrix such as $\Xi$, the integrand in curly brackets is
nonnegative, 
\be
\lsp \mathrm{tr_3} \, \Xi(x) \rsp^2  \  \geq  \  \mathrm{tr_3} \, \lsp 
                    \Xi(x)^2 \rsp  \  , 
\label{trinq} 
\ee
(proof: write the difference in terms of the eigenvalues of $\Xi$) and
the minimum value of zero is attained, apart from the trivial case
$\Xi \E 0$, when $\Xi$ becomes proportional to a projector onto a 
one-dimensional subspace (facts familiar from the properties of the density
matrix). A glance at (\ref{Xicomp}) shows that there is a natural
candidate for a term proportional to a projector -- the $\chi \chi^T$
matrix. The minimum of zero is thus assumed when
\be
\Xi(x) \E \chi(x) \, \chi^T(x) \, ,\qquad \mathrm{i.\,e.}\quad \Psi(x) \E 0 \  , 
\label{Ximin} 
\ee
Recall that in the standard Higgs action, a nontrivial minimum of the
quartic potential at a finite value of the scalar field is ensured through
the addition, by hand, of a ``wrong-sign mass term'', $- \, \mu^2 \, 
\phi^{\dagger} \phi$, which is really just a device for creating a
minimum at a finite and constant value $\phi^{\dagger} \phi \E \frac
{1}{2} \, v^2$, where $v \, \propto \, \mu$. It is interesting to observe
that at least for the U(2) gauge group, an essentially similar structure,
with a minimum at finite field values, is {\em naturally} present
in the $S_4$ term. Also, the term quartic in $\chi$ cancels in the integrand
of eq.~(\ref{S4tr}), so that no mass term can form for the $\chi$s, and the
difference (\ref{trinq}) may be rewritten
\bea
\frac{1}{4} \lcp \lsp \mathrm{tr}_3 \, \Xi(x) \rsp^2  \ - \  \mathrm{tr}_3 \, 
\lsp \Xi(x)^2 \rsp \rcp  
& \E & \mathrm{tr}_3  \lcp \frac{1}{2} \, \Psi^T \, M^2[\chi] \, \Psi \  + \  
\frac{1}{4} \lrp \Psi^T f_a \, \Psi \rrp^{\dagger}  \lrp \Psi^T f_a \, \Psi
\rrp  \rcp   ,  
\label{masstr}                                                             \\
\lcp M^2[\chi] \rcp^{\,a \,b} & \Def & \chi^2(x) \, \delta^{\,a \,b} \ - \   
\chi^a (x) \, \chi^b(x) \  .
\label{masschi} 
\eea
with $f_a$ the adjoint SU(2) generators of eq.~(\ref{fc}) below. Remember
$\Psi$ is real. What is different here is that the minimum field
configuration is not a constant but rather an $x$-dependent quantity,
proportional to the projector
\be
\tilde{\Pi}(x)\ : \qquad \lsp \tilde{\Pi}(x) \rsp^{\,a\,b} \E \frac{ \chi^a (x) \  \chi^b(x) }
{ [\, \chi(x) \,]^2 } \ ; \qquad   \chi(x) \E  \lsp \chi^c (x) \, \chi^c(x)
\rsp^{\frac{1}{2}}
\label{chiproj} 
\ee
onto the $\chi$ direction in isospace, locally at $x$. Thus in general it
minimizes only the $S_4$ term of the action, whereas the standard $\varphi
\E \frac{1}{\sqrt{2}} \, v$ also makes derivative terms vanish, and
thus pushes the entire Higgs-field action into its minimum. We may however
select a constant one among the minimum configurations (\ref{Ximin}), based
on the vacuum expectation values of our scalar fields, which are expected
to settle in the classical potential minimum but are $x$-independent. Thus,
with a hat denoting quantum-field operators,
\be
\chi^a(x) \E  \nu \, u^a \, + \, \chi\,'^a(x) \ ; \qquad  \nu \, u^a \Def \la
0 \lvl \hat \chi^a(x) \rvl 0 \ra  \E  \la 0 \lvl \hat \chi^a(0) \rvl 0 \ra \ 
; \qquad \la 0 \lvl \hat \chi\,'^a(x) \rvl 0 \ra \E 0 \  .  
\label{vacexp} 
\ee
Here $u^a$ is a constant unit isovector, $u^au^a = 1$, and the field
$\chi'^a(x)$, an isovector under global gauging, sweeps the non-constant
remainder of the configurations (\ref{Ximin}).

On the other hand, the field degrees of freedom describing deviations from
the minimum configuration, $\Xi - \Xi_{min} \E \Psi \cdot \Psi^T$, being given
by the elements of the $\Psi$ matrix, are nine in number, rather than the
four of the usual Higgs scenario. The ``mass matrix'' of eq.
(\ref{masschi}) now takes the form,
\be
\lcp M^2[\chi] \rcp^{\,a \,b}  \E  \lsp \nu \, u^c \, + \, \chi\,'^c(x) \rsp^2
\,  \lsp \Id_3 \  - \  \tilde{\Pi}(x) \rsp^{\,a \,b}  \  ,
\label{massmat} 
\ee
proportional to a projector onto a {\em two}-dimensional subspace. The $S_4$
portion of the action therefore splits off a mass term
\be
\frac{1}{2} \, \nu^2 \,g_2^{\,2} \   \mathrm{tr}_3  \lrp \Psi'^T \, , \, 
\Psi' \rrp \   ;  \qquad \quad   \Psi' \Def  \lrp \Id_3 \  - 
\  \tilde{\Pi} \rrp \Psi \  ,
\label{Psimass} 
\ee
with a mass of $m_{scalar}^{\,2} \E  \lrp \nu\,g_2 \rrp^2$, for only 
$3 \times 2 \E 6$ of the nine $\Psi$ fields. The other three,
contained in $\Psi'' \E \tilde{\Pi} \,\Psi$, remain massless, as did the
three $\chi$s. It is these six fields that one might envisage transforming
away by a gauge choice, although at this stage, with only three SU(2)
gauge parameters available, it is not clear how this could work. The
question will, however, come up again in the setting of section 10.

The complete ``all-$Q$'s'' action, upon putting in the $S_2$ pieces from 
eqs.~(\ref{S2diag}) and  (\ref{S2long}) and the $S_3$ term from eqs.~(\ref{S3QQT}) 
and  (\ref{QdQT3}), comes out as 
\bea
S_E \lsp Q,\bar c, c \rsp  - \lrp K, \, Q \rrp   \E 
\frac{1}{2} \,\mathrm{tr}_3 \, \lsp \lrp \partial_{\lambda} \Psi^T \, ,
\, \partial_{\lambda} \Psi \rrp \, + \, (\nu \, g_2)^{\,2} \, \lrp \Psi'^T \, ,
\, \Psi' \rrp  \rsp                          \hspace{1cm}            \non
+ \  \frac{1}{2} \, \lsp  
\  \lrp \partial_{\lambda} \chi\,'^a \, ,\, \partial_{\lambda} \chi\,'^a \rrp
\,  +  \, \lrp \partial_{\lambda} \eta^b \, ,\, \partial_{\lambda} \eta^b \rrp
\  + \  \lrp \partial_{\lambda} \psi \, ,\, \partial_{\lambda} \psi \rrp 
\rsp                                              \hspace{2cm}         \non
+ \ \frac{1}{2}\,g_2  \, \epsilon^{abc} \, \lrp \lsp \, \Psi \drl_{\!\!\!\mu} \, 
\Psi^T \, \rsp^{\,b\,c}  +  \chi\,'^b \drl_{\!\!\!\mu} \, \chi\,'^c 
+  \nu \, \lsp \, u^b \partial_{\mu} \chi\,'^c \, - \, u^c
\partial_{\mu} \chi\,'^b \, \rsp \  , \   \Psi^{\,a\,d} \, n_{\mu}^d \  + \  
\chi^{\,a}  n_{\mu}^4      \rrp                                          \non
+ \ g_2^{\,2}  \, \mathrm{tr}_3  \lcp \, \lrp \Psi'^T \, , \, [\, \nu u^c 
\chi\,'^c \, + \, \frac{1}{2} \chi\,'^c \chi\,'^c \,] \, \Psi' \rrp
\  +  \  \frac{1}{4} \lrp \lsp \Psi^T f_a \, \Psi \rsp^{\dagger}  \  ,
\  \lsp \Psi^T f_a \, \Psi \rsp \rrp \rcp    \hspace{0.5cm}         \non
+ \lcp \frac{1}{2} \, \mathrm{tr}_4 \, 
\lrp \partial_{\mu} Q \, , \, (n_{\,\mu} \cdot n_{\,\nu}^T) \, \partial_{\nu} Q^T 
\rrp \,  +  \, S_{GFG} \lsp Q,c, \bar{c} \rsp \rcp \ - \  \lrp
K , \, Q \rrp .  \qquad \qquad  
\label{actallQ}
\eea
To limit the proliferation of terms, we have refrained from introducing the
splitting  $\Psi \E \Psi' \,+\, \Psi''$ of the $\Psi$ field in the 3rd and
4th lines of eq.~(\ref{actallQ}), as well as in the first term.    

%
%
Although we will be led below to the introduction of a minimal complex
extension of our present real $Q$ fields, this discussion of the $S_4$ term
will be seen to remain valid, provided we replace the transpose $\Psi^T$
everywhere by the hermitean conjugate, $\Psi^{\dagger}$. More important will
be the generalization of $S_4$ obtained in sect.~7 below, which aims at
bringing all {\em four} vector fields into play. 
\vspace{0.4cm}

\section{A covariant derivative}
\setcounter{equation}{0}
%
Conceptually more interesting would be a {\em mixed A-and-Q formulation},
describing vector and scalar field variables on the same level.
For this it is obviously necessary to be able to formulate the interaction of
the scalars with their ``parent'' fields in terms of
(a scalar contraction of) {\em gauge-covariant derivatives}. This,
in particular, is the only known, gauge-invariant way of generating the
seagull-type terms that can provide for vector-mass generation through
vacuum expectations of scalars.

      Now the very concept of a covariant derivative, as commonly understood,
tacitly presupposes {\em homogeneous} gauge/BRS transformation properties
in the fields acted upon, as illustrated by the standard Higgs doublet. We
already emphasized that the scalars considered here exhibit gauge changes
\mbox{(appendix A)} that are at least partly inhomogeneous, but in
building action functionals we require a {\em modified form of covariant
derivative for $Q$ scalars} which still transforms homogeneously so its
contraction with itself will form an invariant. It is interesting that such
a construct,
gauge-transforming homogeneously in spite of the fact that the scalars acted
upon do not, is in fact feasible, and again possesses a relatively simple
interpretation. We approach this construct heuristically, discussing in this
section the pure SU(2) case, and extending it in the following section to
the U(2) context through the introduction of hypercharge.

         We start from the usual form of a covariant derivative for an U(2)
local gauge group,
\be
D_{\mu}  \Def  \Id_4  \partial_{\mu} \, - i g_2 \, A_{\mu}^d F_d
                                      - i g_1 \, B_{\mu}   Y  \  ,
\label{oldcov}
\ee
(the couplings $g_2,\ g_1$ being often denoted alternatively as $ g,\
g^{\prime} $), where the $4 \times 4$ matrices $F^c$ have three-plus-one
partitions,
\bea            
\textnormal{\large $F_c$}   \E  \left( \begin{array}{cccc}
            \multicolumn{3}{c}{ \textnormal{\raisebox{-0.7mm}
                               {{\Large $f_c$}  }} \quad }  & 0        \\
            \multicolumn{4}{c}{ {} }                                   \\
            \multicolumn{3}{c}{ 0 \quad }             & 0        \\ 
                                       \end{array} \right) \  .  
\label{Fc}
\eea  
Here $f^c$ are the hermitean SU(2) generators in the adjoint representation,
\be
\left( f_c \right)^{\,a\,b}  \E  - \, i \, \epsilon_{\,c\,a\,b} \  ,
\label{fc}
\ee
while the hermitean, $4 \times 4$ hypercharge matrix $Y$ should commute with
the three $F^c$'s,
\be
\lsp F_c,\, Y \rsp \E 0 \, .
\label{Ycomm}
\ee
For the moment we do not specify $Y$ further; its form as adapted to our
context -- eq.~(\ref{T4diag}) below -- will emerge only
after we have effected the transformation to a charge-diagonal basis for
our scalars. In this section, since our construction so far has been entirely
in terms of vector gauge fields carrying zero hypercharge, we start by setting
$g_1 \E 0$, which amounts to omitting the $B_{\mu} Y$ term from definition
(\ref{oldcov}) altogether, and to considering
\be
\tilde{D}_{\mu}  \Def  \Id_4  \partial_{\mu} \, - i g_2 \, A_{\mu}^d F_d
\label{tildcov}
\ee
instead. We straightforwardly evaluate, using the properties of the matrices
(\ref{Fc}), the infinitesimal gauge variation of $\tilde{D}_{\mu} Q$ under
the nonhomogeneous gauge variation of $Q$, eqs.~(\ref{Psief}/\ref{chief}).
The result is
\be
\delta_{\theta} \lsp \lrp \tilde{D}_{\mu} Q(x) \rrp^{\,A,\,B} \rsp  \E
\lsp i \, g_2 \, \delta \theta^d(x) \, F_d \rsp^{\,A,\,C} \, \lrp \tilde{D}_{\mu}
Q(x) \rrp^{\,C,\,B} 
\, + \,  \tilde{D}_{\mu}^{\,A,\,C}  \lsp  \lrp \partial_{\nu}
\delta \theta^C(x) \rrp  n_{\nu}^B \rsp  \  ,
\label{deltatild}
\ee
On the r.h.s.\ of eq.~(\ref{deltatild}), the first term gives the infinitesimal
form of an homogeneous transformation law, while the second collects the
contributions arising from the inhomogeneous parts (containing derivatives
of the gauge functions $\delta \theta^C(x)$). We recast the latter term in a
form with the order of the two derivatives interchanged. This leads to
\be
\tilde{D}_{\mu}^{\,A,\,C} \lsp \lrp \partial_{\nu}
\delta \theta^C(x) \rrp  n_{\nu}^B  \rsp  \E  
\delta_{\theta} \lsp \partial_{\nu} \lrp  A_{\mu}^A  n_{\nu}^B \rrp \rsp \, + \,
\lrp i \, g_2 \, \delta \theta^d(x) \, F_d \rrp^{\,A,\,C} \lsp - \partial_{\nu}
\lrp A_{\mu}^C \, n_{\nu}^B \rrp \rsp \  .
\label{inhomterm}
\ee
Using this result in eq.~(\ref{deltatild}), bringing its first term to the
l.h.s.\ of that formula, and applying eq.~(\ref{AQn}), we conclude that the
{\em modified covariant derivative for $Q$ fields without hypercharge},
\bea
\lrp \tilde{\nabla}_{\mu} \, Q \rrp^{\,A,\,B}   & \Def &
\lrp \tilde{D}_{\mu} \, Q \rrp^{\,A,\,B} \, - \, \lrp \partial_{\nu} Q^{\,A,\,C}
     \rrp \lrp n_{\mu} \cdot n_{\nu}^T \rrp^{\,C,\,B}   \  ,    \\
     & \E &  \partial_{\mu} Q^{\,A,\,B} \, - \, \lrp \partial_{\nu} Q^{\,A,\,C}
     \rrp \lrp n_{\mu} \cdot n_{\nu}^T \rrp^{\,C,\,B} \,
     - i g_2 \, A_{\mu}^d (F_d)^{\,A,\,C} \, Q^{\,C,\,B}   
\label{nabtild}
\eea
obeys a purely homogeneous gauge-transformation law,
\be
\delta_{\theta} \lsp \lrp \tilde{\nabla}_{\mu} Q(x) \rrp^{\,A,\,B} \rsp  \E  
\lrp i \, g_2 \, \delta \theta^d(x) \, F_d \rrp^{\,A,\,C} \, \lrp
\tilde{\nabla}_{\mu} Q(x) \rrp^{\,C,\,B}       \  ,
\label{tildhom}
\ee
despite the nonhomogeneous law for $Q$ itself. In eq.~(\ref{nabtild}), one
observes again the occurrence of the tensorial combination (\ref{ncrossn})
of $n$-basis vectors.

     In this no-hypercharge case, the homogeneous-transformation property of
$\tilde{\nabla} Q$ can also be seen in a different way, namely by formally
rewriting it entirely in terms of $A$ fields and $n$ vectors, using again
eq.~(\ref{QprimAn}). One finds for $A \E a \E 1,\, 2,\, 3$,
\be
\lrp \tilde{\nabla}_{\mu} \, Q \rrp^{a\,b}  \E  G_{\mu \,  \nu}^{\,a} \, 
n_{\nu}^b \, ;  \qquad \quad \lrp \tilde{\nabla}_{\mu} \, Q \rrp^{a\,4}  \E 
G_{\mu \,  \nu}^{\,a} \, n_{\nu}^4 \, ,
\label{naba}
\ee
where $G_{\mu \, \nu}^{\,a}$ are the components of the nonabelian field strength,
while for $A \E 4$, 
\be
\lrp \tilde{\nabla}_{\mu} \, Q \rrp^{4\,b}  \E  F_{\mu \,  \nu} \, n_{\nu}^b \, ;
\qquad \quad \lrp \tilde{\nabla}_{\mu} \, Q \rrp^{4\,4}  \E  F_{\mu \, \nu} \, 
n_{\nu}^4 \, , 
\label{nab4}
\ee
with $F_{\mu \, \nu}$ the abelian field strength. Just as the $Q$ fields
themselves can be viewed as scalar projections of the gauge-vector fields,
the components of $\tilde{\nabla} Q$ can thus be interpreted as projections,
with respect to the second of their tensorial indices, of the field strengths
onto the same $n$-vector system. Since the field strengths transform
homogeneously ($F_{\mu \, \nu}$ being a trivial case) while the $n$ system is
invariant, we again conclude that $\tilde{\nabla} Q$, too, transforms
homogeneously.   

        By using completeness of that system in the form
\be
n_{\nu}^b \, n_{\lambda}^b \,+\, n_{\nu}^4 \,n_{\lambda}^4 \E \delta_{\nu \,
\lambda}  \  , 
\label{ncpl}
\ee
we then obtain the gauge-invariant, four-isospace contraction of two such
modified derivatives in the form
\be
\frac{1}{4} \,\mathrm{tr}_4 \, \lsp \lrp \tilde{\nabla}_{\mu} Q(x) \rrp \,
\cdot\, \lrp \tilde{\nabla}_{\mu} Q(x) \rrp^{\dagger} \rsp  
\E  \frac{1}{4} \lrp G_{\mu \, \nu}^{\,a} \, G_{\mu \, \nu}^{\,a}  
\,+\,  F_{\mu \, \nu} \, F_{\mu \, \nu} \rrp  \E   S_{gauge}[\,A\,]  \  .
\label{tiltil}
\ee
In a sense, this contraction is therefore just a fancy rewriting of the
classical gauge action -- a rewriting, though, that captures the interaction
between two different excitations, vector and scalar, of the same
field system.

         It is instructive to digress briefly on how far one can go with
this no-hypercharge scenario in the direction of vector-mass formation.
For this purpose one would declare 
\be
S_{\mathrm{scalar-gauge}}  \E  \frac{1}{4}  \, \int \! d^4 x \,
\mathrm{tr}_4 \, \lsp \lrp \tilde{\nabla}_{\mu} Q(x) \rrp \,\cdot\, \lrp
\tilde{\nabla}_{\mu} Q(x) \rrp^{\dagger} \rsp  
\label{SAQ}
\ee
as the interaction term in a mixed $A$-and-$Q$ formulation of the problem,
as detailed in sect.~8 below. Upon writing this out, one finds, first, a
``kinetic'' term bilinear in the $Q$ fields, which after partial integrations
reproduces the integral of the sum of the two terms encountered already in
eqs.~(\ref{S2diag}) and (\ref{S2long}). Second, the trilinear,
two-$Q$'s-one-$A$ pieces, after some rewriting, turn into the $S_3$ action
term in the ``mixed'' form of eq.~(\ref{S3QQT}). Our focus at this moment is
on, third, the quadrilinear piece, which turns into the $S_4$ action term,
again in its ``mixed'' form as given in eq.~(\ref{S4mix}), and which can be
written,
\bea
S_4 \lsp Q,\,A \rsp \,  &  \E &  \frac{1}{4} \, g_2^{\,2} \, \int \! d^4 x \, 
A_{\mu}^b(x)  \, \left\{ \, \lsp \chi(x)^2  \Id_3  \,-\, \chi(x) \,
\chi^T(x) \rsp  \right.                                                 \non
& \,+\, &  \left. \lsp \, \mathrm{tr}_3 \lrp \Psi(x) \Psi^T(x) \rrp \cdot
\Id_3 \,-\, \Psi(x) \, \Psi^T(x) \rsp \,\right\}^{b\,c} \cdot A_{\mu}^c(x) \ .
\label{S4mass}
\eea
Upon application of eq.~(\ref{vacexp}), this splits off a bare-mass term
\be
\frac{1}{2} \, \lrp \frac{1}{2} \nu^2 \,g_2^{\,2} \rrp \  \lrp A'^T \, , \, 
A' \rrp \   ;  \qquad \quad   A' \Def \lrp \Id_3 \  - 
\  \tilde{\Pi} \rrp \, A \, ,
\label{Amass} 
\ee
which, since $\Id_3 \,-\,\Pi$ projects on two isodimensions, confers a bare
mass of $m_{gauge}^{\,2}  \E  \frac{1}{2} \, \lrp \nu \, g_2 \rrp^2$
on only two of the gauge fields. In this it is phenomenologically deficient
at this stage, and it is not hard to check that the deficiency will not be
cured by the introduction of vector-field mixing. But it is interesting that
upon comparing with (\ref{Psimass}), eq.~(\ref{Amass}) yields a value of
\be
m_{scalar} \   : \   m_{gauge}  \E \sqrt{2}
\label{ratio} 
\ee
for the tree-level mass ratio, for which the standard scenario offers no clue.
This would hardly qualify as an accurate pre- or postdiction, but in view
of the utter simplicity of the reasoning advanced here in its favor, its 
cost-to-benefit ratio may nevertheless be deemed acceptable.

The occurrence of the projector $\Id_3 - \tilde{\Pi}$, as in 
eq.(\ref{massmat}), is of course welcome,
as it allows for the appearance of a massless photon.
But at the same time, the result shows clearly that without hypercharge we
have one interacting vector field too few in the mass game, and this forces
the next point upon us.     
\vspace{0.4cm}

\section{Hypercharge ?}
\setcounter{equation}{0}
%
In order to introduce hypercharge, as a means of bringing in one more
vector field capable of mass formation, it is of course not sufficient to
simply restore the $g_1 \, B_{\mu} \, Y$ term in eq.~(\ref{oldcov}). A
gauge-covariant construct can result only if the $Q$ fields acted upon
are first endowed with hypercharge, in the sense of a {\em homogeneous}
transformation behavior under a local U(1) group, and this is not trivial
 --  we have built the $Q$'s wholly from gauge fields, which carry no
hypercharge. Can we confer upon our scalar fields a quantum number that
their ``parents'' do not have ?

       The answer offered in the following is tentative, and it
has drawbacks that we shall discuss. It does, however, establish the
required transformation behavior, allows for a gauge-covariant extension of
eq.~(\ref{nabtild}), and will presently be seen to go quite some way
in the vector-mixing and mass-formation problems. In the following we proceed
heuristically, arguing that the natural way of embarking on this problem
is to try to further exploit the freedom in $Q$ as encountered in
choosing the $\exp (\, \Omega (x) \, )$  ``phase'' factor in eq.~(\ref{Qpol}).
Instead of taking the phase matrix purely real and orthogonal, we now
allow for a minimal complex extension, writing 
\be
\bar Q \,(x) \E  {\mathrm e}^{ i \, g_4 \, h(x)\, F_4 }  \, Q \,(x)  ,
\label{Qbar}
\ee
where $F_4$ is an hermitean $4 \times 4$ matrix that will turn out to be
related, though not identical, to hypercharge. For the time being, we require
of $F_4$ only that it commute with the other $F$'s:
\be
\lsp F_4 \, , \, F_a \rsp  \E  0  \qquad \lrp a \E 1, \, 2, \, 3 \rrp .
\label{Fcomm}
\ee
Moreover, $h$ denotes the integral
\be
h(x) \E  \int \! d^4 y \, K_{\mu}(x-y) \, B_{\mu} (y) 
\label{hint}
\ee
with a weight function given by
\be
K_{\mu}(x-y)  \E  \frac{x-y}{2 \pi^2 \, \lsp (x-y)^2 \rsp^2}  \ . 
\label{Kafu}
\ee
As a consequence of the relations
\be
K_{\mu}(x-y) \E  -\frac{\partial}{\partial x_{\mu}} \lsp \frac{1}{4 \pi^2 \,
(x-y)^2} \rsp \, ; \qquad \frac{\partial}{\partial x_{\mu}} \, K_{\mu}(x-y) 
\E \lrp -\frac{\partial}{\partial x_{\mu}} \frac{\partial}{\partial x_{\mu}}
\rrp \lsp \frac{1}{4 \pi^2 \, (x-y)^2} \rsp \E \delta^4 (x-y) \  ,
\label{Kprops}
\ee
$h(x)$ has two simple infinitesimal properties:
\bea
\partial_{\mu} \, h(x) & \E &  B_{\mu} (x)    \  ,      
\label{derivh}                                                        \\
\delta_{\theta} \, h(x) & \E &  \delta \theta^4 (x)  \  .
\label{deltah}
\eea
The $\bar Q$ fields are now complex. The use of complex ``dressing
factors'' as in eq.~(\ref{Qbar}) is again not new; similar factors have been
used as far back as the 1930's by Dirac\cite{Dira} to define physical-field
variables for charged fermions. It must be emphasized that the {\em hypercharge
coupling} $g_4$ as introduced through eq.~(\ref{Qbar}) bears no relation to
the constant $g_1$ appearing in the traditional eq.~(\ref{oldcov});
it is one feature in which the $\bar Q$ scalars depart from their vectorial
ancestry. Also, we should remember that eq.~(\ref{Qbar}) does not define
$g_4$ and $F_4$ separately but only their product; some kind of normalization
of $F_4$ will be necessary to calibrate $g_4$.

      We now explore the consequences of postulating the $\bar Q$'s, rather
than the $Q$'s, to be the physical scalar fields (apart perhaps
from a linear mixing that parallels the usual mixing of gauge vectors
into mass and charge eigenstates). From
eq.~(\ref{deltah}) one infers that the $\bar Q$'s now have acquired the
sought-after, homogeneous terms in their gauge responses to
local U(1) transformations with parameter $\theta^4 (x)$ (see appendix A):
\be
\delta_{\theta} \bar Q(x)  \E  {\mathrm e}^{ i \, g_4 \, h(x)\, F_4 } \,
\lsp \delta_{\theta} Q(x) \, + \, i \, g_4 \, F_4 \, \delta \theta^4(x) \,
Q(x) \rsp \  .
\label{varQbar}
\ee
From eqs.~(\ref{derivh}) and (\ref{Fcomm}) it follows that
\be
\lsp \Id_4 \, \partial_{\mu} \, - \, i \, g_4 \, B_{\mu}(x) \, F_4  \rsp
\bar Q (x)  \E  {\mathrm e}^{ i \, g_4 \, h(x)\, F_4 } \, \lsp
\partial_{\mu} \, Q(x) \rsp  \  .
\label{dQbar}
\ee
Using these relations, it is now straightforward to establish that the
quantity
\be
\nabla_{\mu} \bar Q  \Def \, \tilde{D}_{\mu} \bar Q \, - \, \lrp \partial_{\nu}
\, - \, i \, g_4 \, B_{\nu} \, F_4 \rrp  \lrp \bar Q \, n_{\mu} \cdot n_{\nu}^T
\rrp 
\label{nablaD}
\ee
relates to the $\tilde{\nabla}$ of (\ref{nabtild}) by the simple formula
\be
\nabla_{\mu} \bar Q  \E  {\mathrm e}^{ i \, g_4 \, h(x)\, F_4 } \, \lrp
\tilde{\nabla}_{\mu} \, Q \rrp
\label{nabfact}
\ee
and therefore possesses the homogeneous U(2) transformation law
\be
\delta_{\theta} \lsp \nabla_{\mu} \, \bar Q (x) \rsp  \E  i \,
\lsp \, g_2 \, \delta \theta^d(x) \, F^d \, + \, g_4 \, \delta \theta^4(x) \,
F_4 \rsp  \lsp \nabla_{\mu} \, \bar Q (x) \rsp     \  .
\label{nabhom}
\ee
Thus $\nabla_{\mu}$ is our {\em new covariant derivative with hypercharge},
designed to act on complex $\bar Q$ rather than real $Q$ fields. It can be
written more explicitly as
\be
\lrp \nabla_{\mu} \, \bar Q \rrp^{\,A \,B}  \E  \lrp \nabla_{\mu} 
     \rrp^{\,A B, \, C D} \ \bar Q^{\,C \,D}   \  ,
\label{nablaQ}
\ee
where $\nabla_{\mu}$, in addition to being an Euclidean four-vector, now
emerges as a fourth-rank tensor over the four-dimensional isospace:
\bea
\lrp \nabla_{\mu} \rrp^{\,A B, \, C D}   &\Def&  \delta^{\,A \,C} \lsp
\delta^{\,D \,B} \, \partial_{\mu} \, - \, \lrp n_{\mu} \cdot n_{\nu}^T 
\rrp^{\,D \,B} \, \partial_{\nu} \rsp                                   \non
&-& i \lcp g_2 \, \lrp A_{\mu}^e F^e \rrp^{\,A \,C} \, \delta^{\,D \,B}
\, + \, g_4 \, F_4^{\ A\,C} \lsp B_{\mu} \, \delta^{\,D \,B} \, - \, \lrp n_{\mu} 
\cdot n_{\nu}^T \rrp^{\,D \,B} B_{\nu} \rsp  \rcp  \,  .
\label{nabtens}
\eea
The first line here generates in eq.~(\ref{nablaQ}) the part linear in fields
and containing differential operators. The second line produces terms
bilinear in fields; these comprise, as expected, the terms familiar
from the ``old'' eq.~(\ref{oldcov}), but also a new, unfamiliar term
\be
+ \,i \,g_4 \, F_4^{\, A\,C} \,\lrp n_{\mu}\cdot n_{\nu}^T\rrp^{\,D \,B} \,B_{\nu}
\ \bar Q^{\,C \,D} \  .
\label{newB}
\ee
This can be written, using the $A-Q$ relations of eqs.~(\ref{QprimAn}/\ref{AQn}), 
in various equivalent forms, among which we choose, heuristically,
by arguing that a covariant derivative in Euclidean direction $\mu$ should
provide coupling to vector fields $A_{\mu}$, not $A_{\nu}$. Thus the form
\be
- \, i \,g_4 \, A_{\mu}^D  \lsp \lrp - e^{ig_4 h F_4} \, F_4 \rrp^{\,A\,D} \, \lrp
  e^{-ig_4 h F_4} \rrp^{\,4\,C} \, \rsp \, \bar Q^{\,C \,B}  \  
\label{newA}
\ee
is indicated for expression (\ref{newB}) -- incidentally, it is also the
only form having no uncontracted $n$ vectors.

To deal with the square bracket,
we need to say more about the matrix $F_4$. If $F_4$ were restricted to its
upper-left $3 \times 3$ block, requirement (\ref{Fcomm}) would force it to be
a multiple of $\Id_3$, since the three $F_a$'s generate an irreducible
representation (Schur's lemma). Again arguing heuristically, we then note
that the simplest extension of this to four isodimensions, respecting
(\ref{Fcomm}), is to still keep $F_4$ diagonal but to insert a nonzero $(4,4)$
element. Thus we try
\be
F_4 \E \mathrm{diag}  \lcp \sigma_2 , \, \sigma_2 , \,  \sigma_2 , \,
\sigma_4 \rcp \  .
\label{F4diag}
\ee
This turns the square bracket of expression (\ref{newA}) into
\be
\delta^{\,A \,D}  \, \delta^{\,4 \,C} \, \sigma_{(D)} \,
\mathrm{e}^{ig_4 h(x) [\sigma_{(D)} - \sigma_4] } \, ,
\label{squardiag}
\ee
where $(D)$ is not summed over, and has values of $2$ for $D \E 1,2,3,$ and
$4$ for $D \E 4$. The exponent, if nonzero, would produce x-dependent
generator matrices in eq.~(\ref{newA}) through the function $h(x)$, which
in a covariant derivative is excluded. We are forced to assume
\be
\sigma_2 \E \sigma_4 \E  \sigma , \qquad \quad \mathrm{i.\,e.} \quad
F_4 \E \sigma \Id_4 ; \qquad \quad \mathrm{e}^{ig_4 h(x) F_4} \E \lrp
\mathrm{e}^{i g_4 \sigma h(x)} \rrp \cdot \Id_4 \  .
\label{onesigma}
\ee
Thus $F_4$ commutes with everything not only in the three-dimensional but also
in the four-dimensional isospace; the dressing factor reduces to a
number-valued, x-dependent complex phase. This allows us to get rid of a minor
awkward feature of definition (\ref{Qbar}) as combined with eq.~(\ref{Qpol}):
so far, the complex ``phase'' matrix stands to the left, the real one to the
right of the $X^{\frac{1}{2}}$ ``modulus'' matrix, but now we may write
\be
\bar Q (x)   \E   X^{\frac{1}{2}} (x) \, \cdot \, 
\exp \lrp \, {\bar \Omega} (x) \, \rrp  \, ,
\label{Qbarpol}
\ee
with a single, unitary-matrix factor,
\be                            
\exp \lrp \bar \Omega (x) \rrp  \E  {\mathcal P} \, \lcp
\exp \lsp \, \int_{\mathcal{C}(x_0,x)} {\mathcal M}_{\mu} (s) ds_{\mu}
\, + \, i g_4 \sigma \, \int \! d^4 y \, K_{\mu}(x-y) \, B_{\mu} (y) \,
\Id_4 \, \rsp  \rcp \  . 
\label{Obar}
\ee
The ${\mathcal M}_{\mu}$ portion of the exponent is real and antisymmetric,
the $i \, \Id_4$ portion imaginary and symmetric.
With this, our basic factorization of eq.~(\ref{Qdef}) generalizes simply to
\be
X \lrp x \rrp  \E  {\bar Q} \lrp x \rrp \, \cdot \, {\bar Q}^{\dagger} 
\lrp x \rrp \  .
\label{XQbar}
\ee
Returning now to eq.~(\ref{nabtens}) and introducing into it the simplified
form of the term (\ref{newA}), we obtain  
\bea
\lrp \nabla_{\mu} \rrp^{\,A B, \, C D}   & \Def &  \delta^{\,A \,C} \lsp
\delta^{\,D \,B} \, \partial_{\mu} \, - \, \lrp n_{\mu} \cdot n_{\nu}^T 
\rrp^{\,D \,B} \, \partial_{\nu} \rsp                                   \non
& - & \,i\, \tilde{g} \, A_{\mu}^E (x) \, \lrp \tilde{I}_E \rrp^{\,A \,C} \,
\delta^{\,D\,B}\, ,  
\label{Atens}
\eea
featuring the four (now duly constant) $4 \times 4$ matrices
\be
\lrp \tilde{I}_E \rrp^{\,A \,C}  \Def \  c_g\, \delta_{E \, e}  \,
\lrp F_e \rrp^{\,A \,C} \, + \, s_g\sigma \, \lsp \delta_E^{\,4} \,  
\delta^{\,A \,C} \, - \, \delta^A_{\,E} \, \delta^{\,4 \,C} \rsp
\label{Itildmat}
\ee
for $E = 1 \ldots 4$. Here we have parameterized the
couplings $g_2,\,g_4$ in terms of a common coupling $\tilde{g}$ and an
angle $\vartheta_g$ by putting
\bea
g_2 \E \tilde{g} \, c_g \  ; \qquad c_g \, \Def \ \cos \, \vartheta_g \  ,
\label{cg}                                                                \\
g_4 \E \tilde{g} \, s_g \  ; \qquad s_g \, \Def \ \sin \, \vartheta_g \  ,
\label{sg}                                                              \\
\tilde{g} \, \Def \  \sqrt{\,g_2^{\,2} \, + \, g_4^{\,2} \, }.
\label{gtild}
\eea
In more detail, the matrices (\ref{Itildmat}) are,
\bea
\lrp \tilde{I}_d \rrp^{\,A \,C}  &  \E  &  c_g \, \lrp F_d \rrp^{\,A \,C}
\, - \, s_g \sigma \,  \lrp \Theta_d \rrp^{A \, C} 
                     \qquad (D = d = 1 \ldots 3)     \  ,                 
\label{Itild3}                                                            \\
\lrp \tilde{I}_4 \rrp^{\,A \,C}  &  \E  &  s_g \, \lrp F_4 \rrp^{\,A \,C}
\, - \, s_g \sigma \,  \lrp \Theta_4 \rrp^{A \,C}  \E s_g \sigma \cdot
\mathrm{diag} \, \lcp 1, \, 1, \, 1, \, 0 \rcp  \  .
\label{Itild4}
\eea
We have introduced $4 \times 4$ matrices $\Theta_D$, first encountered
in (\ref{squardiag}), with elements,
\be
\lrp \Theta_D \rrp^{A \, C} \E \delta^A_{\,D} \ \delta^{\,4 \, C} \  ,
\qquad \lrp D \E 1\, \ldots \, 4 \rrp \, ,
\label{Thetamat}
\ee
that is, with a single, fourth-column entry. It is these that will turn
our generator algebra into a truly four-dimensional one.

Note that the matrix $\tilde{I}_4$ that has replaced $F_4$,
and now governs the coupling of our scalars to the $A^4_{\mu} = B_{\mu}$ field,
is diagonal. (Up to a factor it is, in fact, the projector onto the
$3$-dimensional isospace). By contrast, the $\tilde{I}_{1,\,2,\,3}$ matrices,
through their $\Theta_d$ terms, are non-hermitean.
Together, these four matrices no more obey the U(2) commutation
relations -- in this sense, the covariance property of eq.~(\ref{nabhom})
is no more manifest. (This comes as no surprise,
since in writing eq.~(\ref{Atens}) we have torn apart the natural unit of
eq.~(\ref{dQbar}) into its two pieces, in order to separate ``kinetic''
and ``interaction'' parts). To be more precise, we calculate the
commutation relations
\bea
\lsp F_a \, , \, F_b \rsp   & \E &  i \, \epsilon_{a\,b\,c} \, F_c  \  ,
\label{FF}                                                       \\
\lsp F_a \, , \, \Theta_b \rsp   & \E &  i \, \epsilon_{a\,b\,c} \, \Theta_c  \ ,
\label{FT}                                                       \\
\lsp \Theta_a \, , \, \Theta_b \rsp   & \E &  0  \  . 
\label{TT}
\eea
These will be recognized as defining the Lie algebra of the group of rigid
motions in three-dimensional Euclidean space, variously denoted as E(3) or
ISO(3), the semi-direct product of the 3-dimensional rotation and translation
groups. (Here, of course, we are talking of {\em iso}space). The effect of
the novel two-fields term (\ref{newB}), in the form (\ref{newA}), has been
to enlarge the U(2) group behind eq.~(\ref{oldcov}) to this isospace Euclidean
group, with the $\Theta_a$ matrices acting as generators of translations in
isospace. It is well known that finite-dimensional representations of this
non-compact group are not unitary, and their generators therefore not
hermitean, which may help remove some of the exotism of the non-hermiticity
of the $\Theta_a$'s, and lend some after-the-fact plausibility to our heuristic
guessing that led to their appearance in (\ref{squardiag}).

A second, isomorphic E(3) group is generated by the $F_c$ together with the
hermitean adjoints of the $\Theta_a$'s:
\be
\lrp \Theta_D^{\dagger} \rrp^{A \, C} \E \delta^{A \,4} \ \delta^C_{\,D} \  ,
\qquad \lrp D \E 1\, \ldots \, 4 \rrp \  ,
\label{Thetadagg}
\ee
which fulfill the same commutation relations. Finally, the matrix $\Theta_4$,
which is hermitean and represents the projector onto the fourth isodimension,
obeys
\be
\lsp \Theta_4 \, , \, \Theta_a \rsp  \E  - \Theta_a \, ; \qquad
\lsp \Theta_4 \, , \, \Theta_b^{\dagger} \rsp \E \Theta_b^{\dagger} \, ,
\label{Theta4comm}
\ee
while commuting, trivially, with the $F_a$.

With the expression resulting from (\ref{Atens}),
\bea
\lrp \nabla_{\mu} \, {\bar Q} \rrp^{\,A \,B}   & \E &   \lsp
\delta^{\,D \,B} \,\delta_{\,\mu \,\nu} \, - \lrp n_{\mu} \cdot n_{\nu}^T \rrp^{\,D \,B}
\rsp \partial_{\nu} {\bar Q}^{\,A \,D}(x)                                   \non
& - & \, i \, \tilde{g} \lsp A^{E}_{\mu}(x) \, \lrp \tilde{I}_E \rrp^{\,A \,C}
\rsp {\bar Q}^{\,C\,B} \  ,  
\label{Anab}
\eea
the contraction-of-covariant-derivatives action with hypercharge,
\be
S_{\mathrm{scalar-gauge}}  \E  \frac{1}{4}  \, \int \! d^4 x \,
\mathrm{tr}_4 \, \lsp \lrp \nabla_{\mu} {\bar Q}(x) \rrp \,\cdot\, \lrp
\nabla_{\mu} {\bar Q}(x) \rrp^{\dagger} \rsp  \, ,
\label{SAQbar}
\ee
finally has terms with zero, one, and two vector fields,
\be
S_{\mathrm{scalar-gauge}} \lsp {\bar Q}, \, A \rsp  \E  
                          C_2 \lsp {\bar Q} \rsp
                      +   C_3 \lsp {\bar Q}, \, A \rsp
                      +   C_4 \lsp {\bar Q}, \, A \rsp   \  ,
\label{ssg}
\ee
where $C_2$ now is a kinetic term for the $\bar Q$ scalars analogous to
(the integral of) the sum of terms (\ref{S2diag}) and (\ref{S2long}),
\be
C_2 \lsp {\bar Q} \rsp  \E  \frac{1}{2}  \, \int \! d^4 x \,
\mathrm{tr}_4 \, \lsp   \lrp \partial_{\mu} {\bar Q}(x) \rrp \,\cdot \,
\lrp \partial_{\mu} {\bar Q}^{\dagger}(x) \rrp  
\, - \, \lrp \partial_{\mu} {\bar Q}(x) \rrp \lrp n_{\mu} \cdot n_{\nu}^T \rrp
\lrp \partial_{\nu} {\bar Q}^{\dagger}(x) \rrp      \rsp      \  ,
\label{C2}
\ee
and $C_3$ is a trilinear interaction of the vector fields with a ``current''
of the $\bar Q$ scalars,
%
%
\bea
C_3 \lsp {\bar Q}, \,A \rsp  \E  \tilde{g}  \int \! d^4 x \, A^{C}_{\mu}(x) 
\, \lrp \frac{-i}{4} \rrp \, \mathrm{tr}_4  \, \lcp \,  +  \, 
{\tilde I}_{C} \lsp {\bar Q}(x) 
\lrp \delta_{\,\mu \,\nu} \Id_4 \, - \, n_{\nu} \cdot n_{\mu}^T \rrp 
\lrp \partial_{\nu} {\bar Q}^{\dagger}(x) \rrp \rsp   \right.           \non
\left.  -  \, ( {\tilde I}_{C} )^{\dagger} \lsp \lrp \partial_{\nu} {\bar Q}(x)
\rrp  \lrp \delta_{\,\mu \,\nu} \Id_4 \, - \, n_{\mu} \cdot n_{\nu}^T \rrp
{\bar Q}^{\dagger}(x) \rsp  \rcp    \  ,
\label{C3}
\eea
a generalization of eq.~(\ref{S3QQT}) to four isodimensions. Finally,
$C_4$ denotes the seagull-type, four-fields term
\be
C_4^{(A)} \lsp {\bar Q}, \, A \rsp   \E  \tilde{g}^2 \int \! d^4 x \,
A^{C}_{\mu}(x)
\lcp \frac{1}{4} \, \mathrm{tr}_4 \, \lsp  ( \tilde{I}_{C} )^{\dagger} \,
\tilde{I}_{D} \, {\bar Q}(x) \, {\bar Q}^{\dagger}(x) \rsp   \rcp
A^{D}_{\mu}(x)  \  ,  
\label{C4AQ}
\ee
which generalizes eqs.~(\ref{S4mix}/\ref{S4mass}) to four isodimensions.
Like the latter, it has an alternative and equivalent form as a
quadrilinear self-interaction of scalars that mirrors eq.~(\ref{S4quart}): 
\be
C_4^{(Q)} \lsp {\bar Q} \rsp   \E  \tilde{g}^2 \int \! d^4 x \, \lrp
{\bar Q}(x) \, {\bar Q}^{\dagger}(x) \rrp^{\,C\,D}
\lcp \frac{1}{4} \, \mathrm{tr}_4 \, \lsp  \tilde{I}_{C} )^{\,\dagger} \,
\tilde{I}_{D} \, {\bar Q}(x) \, {\bar Q}^{\dagger}(x) \rsp \,  \rcp  \ .
\label{C4QQ}
\ee
Eqs.~(\ref{C4AQ}/\ref{C4QQ}) will be the starting point for section 9 below.

          It is clear from the combination of eqs.~(\ref{tiltil}) and
(\ref{nabfact}) that the action of eq.~(\ref{SAQbar}) is still a fancy
rewriting of the classical gauge action. Since the latter ``knows nothing''
about the coupling $g_4$, it follows that (\ref{SAQbar}), contrary to
appearances, is in fact independent of $g_4$. As we shall see in section 9,
this invariance will be broken in the scalar sector by imposing the
condition that the scalar fields carry a truly four-dimensional representation
of the extended isospin -- this condition will be met at a 
nonzero value of $g_4 \sigma / g_2$ (eq.~(\ref{gsigmacond}) below).
In the world of ``intrinsic'' scalars, once they have been endowed with
hypercharge in the way attempted here, there is therefore the possibility of
``spontaneous creation of a coupling''.     
\vspace{0.4cm}

\section{Mixed-variables action}
\setcounter{equation}{0}
%
%
\hspace{15mm} {\bf Enforcing the $A-Q$ Relation}. We now need to address
the fact that use of an action like  (\ref{SAQbar}/\ref{ssg}) makes sense
only in a {\em mixed A-and-Q formulation} of the problem -- which,
incidentally, would also lend itself best to a comparison with the standard
Higgs scenario. Such a formulation involves rewriting the path integral
(\ref{pathint}) as an integral over both vectorand scalar fields, while
imposing the relation between these through delta functionals. In this section
we discuss this rewriting on the simplest level of the original, real $A$ and
$Q$ fields, leaving the modifications for complex fields such as $W^{\pm}$
-- eqs.~(\ref{wplus}/\ref{wminus}) below -- or $\bar Q$ to the reader.
On this level, we have to enforce the relation (\ref{QprimAn}) between the
two kinds of variables by inserting
\be
1 \E  \int \! \mathcal{D}[ Q ] \, \delta \lsp Q \,-\, \, A_{\lambda} \, 
n_{\lambda}  \rsp 
\label{deltafu} 
\ee
into a double path integral:
\bea
Z \lsp J, \, K \rsp  &\E&  \int \! \mathcal{D}[Q] \mathcal{D}[A] 
\mathcal{D}[c,\bar c]\ \, \delta \lsp Q \,-\, 
A_{\lambda} \, n_{\lambda}  \rsp  \,                                      \non 
&\cdot&  \exp \, \lcp -S_{\mathrm{mixed}}[A, Q, \bar c, c] \, + \, (J,A)
\, + \, (K,\,Q) \rcp \, .  
\label{mixint} 
\eea
Here $ S_{\mathrm{mixed}} $ stands for a suitable form of the gauge action,
to be detailed below, which involves both vector and scalar fields.
Exponentiation of the delta functional, in order to convert it into additional
action terms, can be achieved by various techniques. One might insert
\be
\delta \lsp Q \,-\, A_{\lambda} \, n_{\lambda} \rsp  \E  \int \!
\mathcal{D}[R] \, \exp \lcp \, i \, \lrp R, \, \lsp Q - \,
A_{\lambda} n_{\lambda} \rsp   \rrp   \rcp,
\label{fourierdelta}
\ee
as is done in the field-strength formulation of ref.~\cite{Scha}. This
is technically unobjectionable, but as an action term the imaginary exponent
seems somewhat out of place in an Euclidean theory (except when, as in 
ref.~\cite{Scha}, it is immediately absorbed by a Gaussian integration).
Alternatively, one could utilize the limit representation, 
\be
\delta \lsp Q \,-\, A_{\lambda} \, n_{\lambda} \rsp  \E  \lim_{\beta \to 0} \,
\textnormal{const.} \, \exp \lcp \int \! d^{4}x \, \frac{\lrp Q(x) - A_{\lambda}
(x) n_{\lambda} \rrp^2}{2 \beta^2} \, \rcp \  ,
\label{gaussdelta}
\ee
which, before the limit is executed, would resemble the method of Kondo\cite{Kond} 
for introducing new functional variables defined ``on the
Gaussian average''. The (infinite) constant again drops out of the ratio
(\ref{Gfrac}).

           There is yet a third method, more specifically attuned to the
context of a gauge theory, on which we concentrate in the following. It
consists in appealing to a functional theorem that underlies the De Witt\,-
Fadde'ev-Popov quantization procedure. In the formulation of \cite{Wein}, 
and with minimally adapted notation, the theorem runs as follows:

{\bf Theorem}: Let $\phi_n(x)$ be a set of gauge and matter fields and
$\mathcal{D} \phi \E \prod_{n,x} \, d\phi_n(x)$ their functional
volume element. Let $\mathcal{G} [\phi]$ be a functional of the $\phi_n$
satisfying the gauge-invariance condition
\be
\mathcal{G} [ \phi_{\theta} ] \, \mathcal{D} [ \phi_{\theta} ]   \E   
\mathcal{G} [ \phi ] \, \mathcal{D} [ \phi ] \, , 
\label{Gtheo}
\ee
where $\phi_{\theta}$ are the $\phi$ fields locally gauge-transformed with
gauge parameters $\theta^{A}(x)$. Moreover, let $f^{A} [\phi, x)$ be a set of 
gauge-noninvariant functionals of $\phi$ and functions of $x$, and
$B[f[\phi]]$, in turn, a functional of the $f^{A}$. Finally, let $\mathcal F$
denote the continuous matrix of functional derivatives,
\be
{\mathcal F}_{A,x \, ; \, B,y}  \E  \frac{ \delta \, f^{A} [\phi_{\theta}; x) }
{ \delta \, \theta^{B}(y) } \  \Bigr\rvert_{\theta = 0} \   .
\label{funcdiv}
\ee
Then the functional integral           
\be
\mathcal{J}  \E  \int \! \mathcal{D}[ \phi ] \, \mathcal{G} [ \phi ] \, 
\mathcal{B}[f[\phi]] \, \mathrm{Det} \, {\mathcal F} [ \phi ]  
\label{fixint}
\ee
is actually independent (within broad limits) of the noninvariant
functionals $f^{A} [\phi, x)$, depends on the choice of the functional 
$\mathcal{B}[f]$ only through an irrelevant constant factor, and in fact equals
\be
\mathcal{J}  \E  \frac{C}{\Omega} \, \int \! \mathcal{D}[ \phi ] \,
\mathcal{G} [ \phi ]  \  ,
\label{gaugint}
\ee
where $C\, = \, \int \! \mathcal{D}[ f ] \,\mathcal{B}[f]$ is the irrelevant
constant, and where
\be
\Omega  \E   \int \! \mathcal{D}[ \theta ] \, \rho\, [ \theta ]  \qquad
\qquad  \Bigl( \mathcal{D}[ \theta ] \E  \prod_{x \in E^4, A}  d \theta^A (x)
\Bigr)
\label{grpvol}
\ee
is the volume of the gauge group (in other words, $\rho[\theta]$ is the
invariant or Haar measure on the space of group parameters).

      This theorem is usually applied for the purpose of changing between
gauge-fixing schemes, but it can also be used as a method of enforcing
relations between functional variables. In the present context, it evidently
cannot be applied to the $\mathcal{D}[ Q ] \, \mathcal{D} [ A, c, \bar c ]$
integration of eq.~(\ref{mixint}) as a whole, since the obvious choice for
the ``initial'' functional $f$,
\be
(f^{(0)})_{\mu}^{\,C} \, [ \left. Q, A; x ) \right. \E  A^C_{\mu} \,-\,
Q^{C\,D} \, n^D_{\mu}
\label{f0func}
\ee
would, in the presence of the ``initial'' $\mathcal{B}$ functional
\be
\mathcal{B}^{(0)} \lsp f^{(0)} \lsp A \rsp \rsp  \E  \delta \lsp f^{(0)} \rsp  \E
\prod_{\mu = 1}^4 \, \lcp \prod_{C = 1}^4 \, \prod_{x \in E^4} \delta \lsp
(f^{(0)})_{\mu}^{\,C} [ \left. Q, A; x ) \right. \rsp  \rcp  \E
\prod_{\mu = 1}^4 \, \mathcal{B}^{(0)}_{\mu} \lsp f^{(0)}_{\mu} \rsp \  ,  
\label{B0func}
\ee
be effectively gauge invariant under combined gauge transformation
of {\em both} the $A$ and $Q$ fields (the kernel ${\mathcal F}$ of 
eq.~(\ref{funcdiv}) would vanish). It can, however, be applied to the 
``inner'' $A$ integration at any {\em fixed} configuration $Q$ of the 
``outer'' integration of (\ref{mixint}), where reference to the fixed $Q$
breaks invariance under
gauging of $A$ alone. We also require a slight modification of the above
theorem, since the $f^{(0)}$ of (\ref{f0func}) now also carries a spacetime
index $\mu$. Here we observe that the gauge transformations on $A$ fields
do not mix the spacetime indices, and that the $\mathcal{B}^{(0)}$ functional
of eq.~(\ref{B0func}), as well as the functional volume element
$\mathcal{D} [A] \E \prod_{\mu = 1}^4 \,\mathcal{D}[A_{\mu}]$,
have a product structure with respect to $\mu$. Therefore
the theorem can be applied consecutively to each of the four nested path
integrations on $A_{\mu}$. For this, however, it is necessary at each step
to first multiply and divide by the corresponding functional determinant
$\mathrm{Det} \, {\mathcal F}^{(0)}_{\,\mu}$; the inverse determinant, which
in the presence of the $\delta$ functionals effectively depends only on $Q$,
can be shifted into the ``outer'' $Q$ integration. The theorem then permits,
for one $\mu$ at a time, the replacements
\be
(f^{(0)})_{\mu}^{\,C} \ \longrightarrow \  f_{\mu}^{\,C} \, ,  \qquad
\mathcal{B}^{(0)}_{\,\mu} \lsp f^{(0)} \rsp \ \longrightarrow \  \mathcal{B}_{\mu}
\lsp f_{\mu} \rsp \, ,  \qquad  \textnormal{Det} \, {\mathcal F}^{(0)}_{\,\mu}
\ \longrightarrow \  \textnormal{Det} \,{\mathcal F}_{\mu}
\label{apptheo}
\ee
inside the $A$ integral, with wide freedom in the choice of the ``final''
$f_{\mu}$ and $B_{\mu}$ functionals.

      For the functional-derivative kernel ${\mathcal F}^{(0)}_{\,\mu} \E \delta
f^{(0)} \, / \, \delta \theta$, using the notation of eq.~(\ref{delbar}) of
appendix A and the presence of the delta functional $\mathcal{B}^{(0)}_{\,\mu}$,
we obtain 
\begin{gather}
({\mathcal F}^{(0)}_{\,\mu})_{B,x \, ; \, C,y}  \E  \frac{ \delta \, (f^{(0)})_{\,\mu})
^{B} [\,A_{\theta}; x)}{ \delta \, \theta^{C}(y) } \  \Bigr\rvert_{\theta = 0,
A_{\mu} \rightarrow Qn_{\mu} }                                            \non
\E \lcp \lsp \frac{\partial}{\partial x_{\mu}} \, \delta^4 (x-y)
\rsp \delta^{B \, C} \,  - \, g_2 \, \delta^4 (x-y) \, \epsilon
^{B\,C\,d} \, Q^{d\,E}(y) n_{\mu}^{E} \rcp  \  .
\label{F0kern}
\end{gather}
The inverse determinants of these kernels, having accumulated outside
the $A$ integration, form a product-inverse determinant that can be
represented by a path integral over {\em bosonic} ghost fields:
\bea
\lsp \mathrm{Det}\,{\mathcal F}^{(0)} \rsp^{(-1)}  \E  \prod_{\mu = 1}^4 \,
\lsp \mathrm{Det}\,( {\mathcal F}^{(0)}_{\,\mu} )^{(-1)} \rsp              \non
\E  \text{const.} \, \int \! \mathcal{D}[b,\,{\bar b}] \  
\exp \,\lsp - \sum_{\mu=1}^4 \, \lrp {\bar b}^A_{\,\mu} \, , \, 
( {\mathcal F}^{(0)}_{\,\mu})_{A\,B} \, b^A_{\,\mu} \rrp \rsp\  ,
\label{beeghosts}
\eea
with volume element
\be
\mathcal{D}[b,\,\bar b]  \Def  \prod_{\mu = 1}^4 \, \prod_{A = 1}^4 
\lcp \mathcal{D}[b^A_{\,\mu}] \mathcal{D}[{\bar b}^A_{\,\mu}] \rcp \ .
\label{Debee}
\ee
Since ${\mathcal F}^{(0)}$ has unit mass dimension, the $b$'s and $\bar b$'s
are bosonic variables with the unconventional mass dimension of $3/2$.

        For the ``final'' functionals of the process, which should provide
for the desired exponentiation of the delta functionals, we follow closely
the gauge-fixing examples by choosing
\begin{gather}
\mathcal{B} \lsp f \rsp  \E  \prod_{\mu = 1}^4 \, \mathcal{B}_{\mu} \lsp
f_{\mu}[A] \rsp \E
\exp \, \lsp - \frac{1}{2 \zeta} \, \lrp f_{\mu}^{\,C}[A]\,,\,f_{\mu}^{\,C}[A] 
\rrp \rsp   \  ;
\label{Bfunc}                                                      \\    
f_{\mu}^{\,C}[A]  \E  \partial_{\mu}\,g^C[A] \, ; \qquad g^C[A] \E 
A_{\nu}^C n_{\nu}^4 \, - \, Q^{C\,4} \  ,
\label{ffunc}
\end{gather}
with $\zeta$ a dimensionless, real parameter. Then $\mathcal{B}[f]$ is
accompanied by the product of the determinants of the four
functional-derivative kernels 
\begin{gather}
({\mathcal F}_{\,\mu})_{B,x \, ; \, C,y}  \E  \frac{ \delta \,f_{\mu}^{\,B} 
[\,A_{\theta}; x) }{ \delta \, \theta^{C}(y) } \  \Bigr\rvert_{\theta = 0}  \non
\E  \lcp \lsp \frac{\partial^2}{\partial x_{\mu} \, \partial 
x_{\nu}} \, \delta^4 (x-y) \rsp  \delta^{B \, C} \,  
- \, g_2 \, \lsp \frac{\partial}{\partial x_{\mu}} \, \delta^4 (x-y) \rsp \,
\epsilon^{B\,C\,d} \, A^d_{\nu} (y)  \rcp \, n^4_{\nu}  \  .
\label{Fkern}
\end{gather}   
Since this can be written as
\begin{gather}
({\mathcal F}_{\,\mu})_{B,x \, ; \, C,y}  \E  \int \! d^4 z \,
\lsp \frac{\partial}{\partial x_{\mu}}\,\delta^4 (x-z) \,n^4_{\nu} \rsp \non
\times \lcp \lsp \frac{\partial}{\partial z_{\nu}} \, \delta^4 (z-y)
\rsp \delta^{B \, C} \, - \, g_2 \, \delta^4 (z-y) \, \epsilon
^{B\,C\,d} \, A^d_{\nu}(y)  \rcp \  , 
\label{Ffacto}
\end{gather}
and since the determinant of the square-bracketed kernel in the first line
is a field-independent factor that cancels in the ratio (\ref{Gfrac}), we may
as well replace                 
\begin{gather}
({\mathcal F}_{\,\mu})_{B,x \, ; \, C,y}     \,\longrightarrow\,    
({\mathcal F}^{\text{(red)}}_{\,\mu})_{B,x \, ; \, C,y}[A]              \non  
\E  \lcp \lsp \frac{\partial}{\partial x_{\mu}} \,\delta^4 (x-y)
\rsp \delta^{B \, C} \, - \, g_2 \, \delta^4 (x-y) \,
\epsilon^{B\,C\,d} \, A^d_{\mu}(y)  \rcp \  ,
\label{Fredu}
\end{gather}
a first-order differential kernel almost identical to 
${\mathcal F}^{(0)}_{\,\mu}$, but depending on $A$ rather than $Q$. (Remember
the delta functional, at this stage, is no more available). Then
\be
\mathrm{Det} \,{\mathcal F}^{\text{(red)}} \E \prod_{\mu = 1}^4 \, \mathrm{Det}\, 
{\mathcal F}^{\text{(red)}}_{\mu}
\label{Freddet}
\ee
can be represented as an integral over additional Grassmannian
ghost fields, but one may of course also opt for a purely bosonic-ghosts
representation of the quotient kernel $\lcp ({\mathcal F}^{(0)}[Q]) \,
({\mathcal F}^{\text{(red)}}[A])^{-1} \rcp^{-1}$. We see no necessity to spell
these alternatives out in detail. The point of this exercise has been to 
show that the task of enforcing the $Q-A$ relation in the double path integral
can be shifted onto a purely {\em kinetic} addition, eqs.~(\ref{Bfunc})/(\ref{ffunc}), 
to the $A-Q$ action, at the expense of a substantial increase
in the number of ghost-field integrations.

         The still unbroken invariance of the {\em combined} $A-Q$ path
integrand under simultaneous gauge transformation of both $A$'s and $Q$'s
should, at least for a perturbative treatment, be dealt with by
standard schemes of gauge fixing, such as those implemented by the term
$S_{GFG}$ and the concomitant Grassmann ghosts $\bar c, c$ in eq.~(\ref{mixint}).

       {\bf A-Q Action functional}. We must also reorganize the action to
describe {\em both} $A's$ and $Q's$ and their interplay. At first sight, 
it would seem possible to work with an arbitrary mixture 
\be
S_E \E \lrp 1 \, - \, \alpha \rrp \, S_E[\,A,\bar c, c; \,g_2, \,)
               \ + \  \alpha      \, S_E[\,Q, \,A, \, \bar c, c; \,g_2, \,)
\label{actmix} 
\ee
(in self-explanatory notation), with $\alpha$ real. But then we have to
provide normalized kinetic energies $S_2$ for both sets of variables, which
calls for a rescaling of the latter of the form
\be
A \, \To A^{\prime} \E \sqrt{1-\alpha} \  A \  ; \qquad
Q \, \To Q^{\prime} \E \sqrt{\alpha}   \  Q \  .
\label{rescal} 
\ee
(The infinite Jacobian for this rescaling is again a constant that drops out 
of the ratio (\ref{Gfrac})). However, we emphasized already in connection
with condition (\ref{diffcond}) that scaling its two sides differently would
invalidate the derivation of eq.~(\ref{diffeq}) and the entire subsequent
construction of $Q$ right down to eq.~(\ref{QprimAn}). If we wish to hold on
to that construction, then apart from the trivial cases $\alpha \E 0$ and
$\alpha \E 1$, the choice $\alpha \E 1 \,-\, \alpha \E \frac{1}{2}$ is the
only one admissible. 

           The ensuing rescaling with $\frac{1}{\sqrt{2}}$
produces a sum of normalized kinetic terms as in (\ref{S2}) for $A^{\prime}$
and in (\ref{S2diag}/\ref{S2long}) for $Q^{\prime}$. In the tri- and
quadrilinear terms, it produces in addition a rescaling $\tilde{g} \, \To 
\tilde{g} \,\sqrt{2}$ of the coupling constant (\ref{gtild}).
It also extends to the ghost fields and coupling constant in $S_{GFG}$.
Since the quartic term $C_4$ of the $A-Q$ action (\ref{SAQbar}) is needed
both as a seagull-type vector-scalar interaction, and as a self-interaction
of the scalars, and since with the rescaling
of $\tilde{g}$ it appears effectively with a factor of two, we may repeat the
equal-weights decomposition here, writing, with couplings now indicated as
additional arguments,
\be
C_4 \lsp {\bar Q},\,A; \, 2 \tilde{g}^2 \rrp  \E
C_4^{(A)} \lsp {\bar Q},\,A;\, \tilde{g}^2 \rrp  \,+\, C_4^{(Q)} \lsp {\bar Q};
\, \tilde{g}^2 \rrp \  . 
\label{C4split}
\ee
Each of the latter two terms carries the unrescaled coupling $\tilde{g}^2$
whereas the other terms, such as $C_3$ or the pure-$A$ part of the action,
appear with rescaled couplings. (That the total is still gauge invariant
is then no more immediately visible). The schematic structure of the
action is
\bea
S_{\text{mixed}} \lsp A,\,\bar{Q},\,\bar c, c;\,\rrp \E
S_{\text{gauge}} [\,A, \,\bar c, c; \,g_2 \sqrt{2} \,) \,+\, S_{GFG}[\,A, \,
\bar c, c;\,g_2 \sqrt{2} \,)                          \non
\,+\, C_2 [ \bar{Q} ] \,+\, C_3 [\,\bar{Q},\,A; \,\tilde{g} \sqrt{2} \,) \,
  +\, C_4^{(A)} [\, {\bar Q},\,A;\,\tilde{g} )  \,+\, C_4^{(Q)} [\,{\bar Q};
\,\tilde{g} ) \  .
\label{totact}
\eea
To this should be added the exponent of expression (\ref{Bfunc}),
\be
\frac{1}{2 \zeta}  \lrp f_{\mu}^{\,C}[A]\,,\,f_{\mu}^{\,C}[A] \rrp
\E \frac{1}{2 \zeta}  \lrp (\,A_{\mu} n^4_{\mu}\,-\,\phi_C \,) \, , 
\,(-\partial_{\lambda} \partial_{\lambda} )\,(\,A_{\nu} n^4_{\nu}
\,-\,\phi_C \,)  \rrp  \  ,
\label{Bfuncexp}
\ee
as well as the exponents and extra ghost integrations representing
the functional determinants \linebreak 
$[\mathrm{Det}\,{\mathcal F}^{(0)} ]^{-1}\,[ \mathrm{Det}\,{\mathcal F} ]$,
as discussed above. The fields entering here are really the primed ones of
eq.~(\ref{rescal}), but since they are only path-integration variables, we
may as well drop the primes.
\vspace{0.5cm}

\section{Charge-diagonal representation}
\setcounter{equation}{0}
%
In order to decide which of the scalar fields are eligible for VEV
formation, we must first of all select those with zero electric charge,
i.\,e.\ linear combinations of the Q-isomatrix components that belong to
eigenvalue zero of the electric-charge matrix (in units of the positron
charge $e$),
\be
\frac{1}{e} E \E T_3 + Y  \  , 
\label{electric}
\ee
where $T_3$ and $Y$ denote, respectively, the diagonalized forms of the
third component of isospin and of hypercharge. But at this moment, it is
not yet clear what the isospin and hypercharge matrices are in a truly
four-dimensional representation of U(2). The $F_a$ matrices of
eq.~(\ref{Fc}), being just trivial extensions of the three-dimensional
generators, are not satisfactory. The $\tilde{I}_A$ matrices of
eq.~(\ref{Itildmat}) do not qualify, as the first three lack hermiticity,
and the fourth does not commute with them. We have once more to rely
on heuristic procedure.

For a four-dimensional representation of the isospin generators, the simplest
trial choice that suggests itself is to take the {\em hermitean parts} of the
$\tilde{I}_a$. It is not the least among the little surprises offered by
this study that this seemingly simplistic concept actually works. Since
the calculations involved are all standard matrix algebra, we proceed
straight to a statement of the results.

After a further rescaling of the coupling according to
\be
\tilde{g} \ \rightarrow \  2 \, c_g \, \tilde{g} \E 2g_2 \  ,
\label{geescal}
\ee
which restores the original, nonabelian coupling $g_2$ but with a factor of 2,
and a corresponding rescaling of the $\tilde{I}_A$ matrices,
\bea
I_C \, \Def \  &=& \ \frac{1}{2c_g} \, \tilde{I}_C  \  ;
\label{Iiscal}                                                          \\
\lrp I_C \rrp^{\,A \,B} \, &=&  \frac{1}{2}\, \delta_{C \, c}  \,
\lrp F_c \rrp^{\,A \,B} \, + \, \frac{s_g\sigma}{2c_g} \, \lsp \delta_C^{\,4} \,  
\delta^{\,A \,B} \, - \, \delta^A_{\,C} \, \delta^{\,4 \,B} \rsp   \  , 
\label{Imat}
\eea
so that $\tilde{g}\,\tilde{I} \E 2g_2 \, I$, the hermitean parts of the first
three matrices,
\be
R_a \, \Def \  \frac{1}{2} \lrp I_a \, + \, I_a^{\dagger} \rrp
\E \frac{1}{2}\, F_a \, - \, \frac{s_g\sigma}{4c_g} \,
                 \lrp \Theta_a \, + \, \Theta_a^{\dagger} \rrp  \  ,
\label{Rmats}
\ee
together fulfill the SU(2) commutation relations,
\be
\lsp R_a \, , \, R_b \rsp   \E   i \, \epsilon_{a\,b\,c} \, R_c  \  ,
\label{Rcomm}
\ee
{\em provided} the parameters $\theta_g$ and $\sigma$ occurring in them
obey the condition   
\be
\frac{s_g\sigma}{2c_g}  \E  1 \  ,  \qquad \textnormal{or} \quad
g_4 \sigma \E 2g_2 \  .
\label{gsigmacond}
\ee
This relation simplifies the $I_C$ matrices, as well as the $R_a$ themselves:
\bea
\lrp I_C \rrp^{\,A \,B} \  &=& \   \frac{1}{2}\, \delta_{C \, c}  \,
\lrp F_c \rrp^{\,A \,B} \, + \, \lsp \delta_C^{\,4} \,  
\delta^{\,A \,B} \, - \, \delta^A_{\,C} \, \delta^{\,4 \,B} \rsp   \  , 
\label{Isimp}                                                           \\
R_a \  &=& \  \frac{1}{2} \lsp F_a \, - \,
              \lrp \Theta_a \, + \, \Theta_a^{\dagger} \rrp \rsp  \  .
\label{Rsimp}
\eea
The latter, in addition to being hermitean, are traceless and normalized,
\be
tr_4 \, R_a \E 0 \, ;  \qquad \qquad  \parallel \, R_a \, \parallel \E 1 \  , 
\label{Rtrnorm}
\ee
where the usual matrix norm has been employed:
\be
\parallel \, M  \, \parallel^{\,2}  \E  \lrp M\, , \, M \rrp \,  ;   \qquad
\qquad \lrp M_1\, , \, M_2 \rrp  \Def  tr_4 \lrp M_1^{\,\dagger} \, M_2 \rrp \  . 
\label{Matnorm}
\ee
It is remarkable that condition (\ref{gsigmacond}) -- that is, the condition
for the $R_a$ matrices to form a four-dimensional set of generators for the
isospin-SU(2) group -- has fixed the $I_C$ and their hermitean parts $R_c$
{\em absolutely}, with no reference anymore to $c_g, s_g$, or $s_g \sigma$.
The conclusions we draw in the next section from the (rescaled) action terms
(\ref{C4AQ}) and (\ref{C4QQ}), concerning VEV's, masses, and mixing, will
therefore not depend on either of these quantities.

It is then possible, but neither necessary nor of much consequence, to force
the rescaling factor $2c_g$ to be unity, in which case
\be
s_g \E \frac{\sqrt{3}}{2} \  ; \qquad c_g \E \frac{1}{2} \  ,
\label{twocgis1}
\ee
and the scale $\sigma$ of the $F_4$ matrix is then fixed, by condition
(\ref{gsigmacond}), at $\sigma \E 2 \, / \,\sqrt{3}$. But it would just
as well be possible to work the other way around and normalize $F_4$
by setting
\be
\sigma \E 1 \  ;  \qquad \mathrm{i.\,e.}  \qquad F_4 \E \Id_4 \  ,
\label{sigun}
\ee
which, again by (\ref{gsigmacond}), would entail $s_g/c_g \E 2$, or
\be
s_g \E \frac{2}{\sqrt{5}} \  ; \qquad c_g \E \frac{1}{\sqrt{5}} \  .
\label{sgeecgee}
\ee
It is obvious that neither of these choices have anything to do with the
usual electroweak mixing angle.

To extend the four-dimensional representation to the full U(2) group,
we would like to find a hypercharge-generator matrix $R_4$ to match the $R_a$,
but since this can be done properly only after diagonalizing $R_3$, we defer
this point for a moment; the result in eq.~(\ref{R4}) below will be seen
to have all the required properties,
\be
\lsp R_4 \, , \, R_a \rsp   \E   0  \ ; \qquad tr_4 \, R_4 \E  0  \ ; \qquad 
\parallel \, R_4  \, \parallel  \E  1  \  ,
\label{R4props}
\ee
for a consistent and uniformly normalized set of U(2) generators.

The ``leftovers'' $L_A$ in the decomposition 
\be
I_C \E R_C \, + \, L_C  \qquad \qquad ( C \E 1 \, \ldots \, 4 \, )
\label{RLdec}
\ee
are, for $C \E c \E 1 \ldots 3$, the antihermitean parts
\be
L_c \, \Def \  \frac{1}{2} \lrp I_c \, - \, I_c^{\dagger} \rrp \E - \,
\frac{1}{2} \, \lrp \Theta_c \, - \, \Theta_c^{\dagger} \rrp  \E - \,
L_c^{\dagger}  \  ,
\label{Lmats}
\ee
which are real and antisymmetric. For $C=4$, the determination of $L_4$
will once more have to wait for a moment, until the correct choice of $R_4$
has been identified. The matrices (\ref{Lmats}) do not generate a group,
as is evident from their commutation relations written in the form
\be
\lsp -2L_a \, , \, -2L_b \rsp   \E  - \, i \, \epsilon_{a\,b\,c} \, F_c  \  .
\label{LLcomm}
\ee
Taken together with the relations
\be
\lsp F_c \, , \, -2L_d \rsp   \E   i \, \epsilon_{c\,d\,e} \, (-2L_e)  \  ,
\label{FLcomm}
\ee
they will be recognized as those of the boost transformations in the
homogeneous Lorentz group, but associated with the ``old'' rotation
generators $F_c$, rather than with the ``new'' $R_a$. Since there is no
obvious physical meaning to the notion of a boost in isospace, these
observations remain purely formal.

The four-dimensional U(2) representation generated by the $R_A$ is
reducible. The {\em reducing transformation}
\be
T_A  \E   U \, R_A \, U^{\dagger} \  , \qquad \quad ( \, A \E 1 \ldots 4 \, )\  ,
\label{Ttrans}
\ee
which converts all $R_A$'s to a $2 \oplus 2$ block-diagonal form, while
making $T_3$ diagonal, is effected by the unitary matrix
\be
U \E \frac{1}{\sqrt{2}} \, \cdot \,
\begin{pmatrix}
  -1 & i & 0 & 0  \\
  0  & 0 & 1 & 1  \\
  0  & 0 & 1 & -1 \\
  1  & i & 0 & 0
\end{pmatrix}  
\label{Umat}
\ee
with inverse,
\be
U^{\dagger} \E \frac{1}{\sqrt{2}} \, \cdot \,
\begin{pmatrix}
  -1 & 0 & 0  & 1   \\
  -i & 0 & 0  & -i  \\
  0  & 1 & 1  & 0   \\
  0  & 1 & -1 & 0
\end{pmatrix}  \  .
\label{Udagmat}
\ee
(No danger of confusion with the matrix function (\ref{Umatfunc}. The first
three $T$'s then assume the $2 \oplus 2$-block form
\be
T_a \E
\begin{pmatrix}
 \frac{1}{2} \, \tau_a   &    {\textbf {\large 0}}   \\
 {\textbf {\large 0}}    &    \frac{1}{2} \, \tau_a
\end{pmatrix}  
 ( \, a \E 1 \ldots 3 \, )\  ,
\label{Tmat}
\ee
where $\tau_a$ are the Pauli matrices, so that in particular
\be
T_3  \E  \mathrm{diag} \lcp + \frac{1}{2} \  , \, - \frac{1}{2},
                       \    + \frac{1}{2} \  , \, - \frac{1}{2} \rcp
\label{T3diag}
\ee
is diagonal. In this new representation, the hypercharge and electric-charge
matrices, too, should then be diagonal. From the irreducibility of the
fundamental representation generated by the $\tau_a$, it is clear that for
the hypercharge matrix $T_4 \E Y$ to commute with the $T_a$'s requires it to
be a multiple of unity in each block:
\be
T_4 \E Y \E
\begin{pmatrix}
     c_{+} \, \Id_2         &  {\textbf {\large 0}}   \\
 {\textbf {\large 0}}  &      c_{-} \, \Id_2 
\end{pmatrix}  
\qquad  ( \, a \E 1 \ldots 3 \, ) \  .
\label{T4}
\ee
For more insight we turn to the transformation of the scalar fields. We
would expect the isomatrix $Q(x)$ to transform into $\Phi(x) \E U \, Q(x) \,
U^{\dagger}$, which is given in eq.~(\ref{exmatPhi}) of the appendix, but since
only the product $\Phi\,\Phi^{\dagger} \E (UQ)(UQ)^{\dagger}$ appears in the
$C_4$ action terms, and $\Phi \, \partial \,\Phi^{\dagger}$ in $C_3$, it is
sufficient in practice to consider the simpler quantity $U\,Q(x)$. This
also seems more consistent with the fact that the fourth column of $Q$, being
the four-isovector $q$ of (\ref{smallq}) as appearing in the
decomposition (\ref{Xcomp}), would naturally transform into
%
%
\be
\phi(x) \Def \, U \, q(x)  \E
\begin{pmatrix}
  \phi_1 \E \frac{1}{\sqrt{2}} \, \lrp -\chi_1 \,+\,i\,\chi_2 \rrp           \\
  \hspace{-15pt} \phi_2 \E \frac{1}{\sqrt{2}} \, \lrp \chi_3 \,+\, \psi \rrp \\
  \hspace{-15pt} \phi_3 \E \frac{1}{\sqrt{2}} \, \lrp \chi_3 \,-\, \psi \rrp \\
  \phi_4 \E \frac{1}{\sqrt{2}} \, \lrp +\chi_1 \,+\,i\,\chi_2 \rrp
\end{pmatrix} \ .
\label{qtrans}
\ee
This says it all, for every column of $U\,Q$. (Moreover, the fields in the
remainder matrix $U\,P(x)$ will be easier to interpret physically in section
10 below). The electric
charges here are self-suggestive: $\phi_{2,\,3}$ are real scalar
fields and should therefore be electrically neutral. That is, the 2nd and
3rd eigenvalues in the diagonal matrix (\ref{electric}) should vanish.
Taking this together with eq.~(\ref{T3diag}) we conclude that $c_{\pm} \E
\pm \frac{1}{2}$. This in turn gives $\pm 1$ for the 1st and 4th charge
eigenvalues, which fits the interpretation of $\phi_{1,\,4}$ as a  charge-
conjugate pair of fields with integer charges $\pm 1$. It is remarkable
how naturally these assignments, which are essentially those of the standard
Higgs sector, flow from the reduction of the $R$ representation. We now have
\bea
Y   \E   T_4  \, &=& \,  \mathrm{diag} \lcp + \frac{1}{2}, \ + \frac{1}{2},
                         \, - \frac{1}{2}  \, - \frac{1}{2} \rcp \  ,   
\label{T4diag}                                                        \\
\frac{1}{e} E \, &=& \,  \mathrm{diag} \lcp + 1, \   0, \  0, \  - 1 \rcp .
\label{Eldiag}
\eea
The four $T_A$'s again form an hermitean, traceless, and uniformly normalized
set of generators for what we now recognize as the {\em direct sum of two
fundamental representations} of U(2), distinguished only by their $Y$
eigenvalues of $\pm \frac{1}{2}$.

We may now go back and determine the $R_4$ hypercharge generator of the
$R$ representation by applying the $U$ transform backwards:
\be
R_4  \E   U^{\dagger} \, T_4 \, U  \E 
\begin{pmatrix}
\frac{1}{2} \, \tau_2    &  {\textbf {\large 0}}   \\
 {\textbf {\large 0}}    &  \frac{1}{2} \, \tau_1
\end{pmatrix}  \  .  
\label{R4}
\ee
Its commutativity with $R_3$ is obvious, since (\ref{Rsimp}) for $a=3$
has the block-diagonal form
\be
R_3  \E  
\begin{pmatrix}
 \frac{1}{2} \, \tau_2   &  {\textbf {\large 0}}   \\
 {\textbf {\large 0}}    &  - \frac{1}{2} \, \tau_1
\end{pmatrix}  \  ,
\label{R3}
\ee
but it also extends, less obviously, to $R_{1,\,2}$. (From the fact that
$R_3 \, + \, R_4$, the not-yet-diagonal precursor of the electric charge
(\ref{electric}), has a vanishing lower-right block, we might have already
concluded that there must be two vanishing scalar-charge eigenvalues). The
corresponding matrix $L_4$ follows from formulas (\ref{RLdec})
and (\ref{Isimp}):
\be
L_4  \E  
\begin{pmatrix}
\Id_2 \, - \, \frac{1}{2} \, \tau_2    &  {\textbf {\large 0}}      \\
{\textbf {\large 0}}    &   \pi_{+} \, - \, \frac{1}{2} \, \tau_1
\end{pmatrix}   \  \E  L_4^{\,\dagger} \  ,                                    
\label{L4}
\ee
where $\pi_{+} \E \frac{1}{2} \, ( \Id_2 \,+\, \tau_3 )$ is the projector
onto the upper element of a 2-isospinor. Like $R_4$, the matrix
is block diagonal; its matrix norm is $\parallel L_4 \parallel \E 2$. We do
not dwell on its various commutation relations, since these do not lead
to a closed algebra until the level of the general linear group in four
dimensions, $GL(4)$ -- a framework too wide to be of interest
in our context.

While the precise group-theoretic status of the $L_A$ matrices remains obscure,
their contribution will still be important when extracting masses from the $C_4$
action terms, since there it is not only the $R_C$ parts but the complete $I_C$
matrices that come into play, albeit in the hermitean bilinear combination
of eq.~(\ref{Mmat}) below.
\vspace{0.5cm}

\section{VEVs, Masses, and Mixing}
\setcounter{equation}{0}
%
The question of mass formation, for both the vector and scalar fields, now
poses itself anew, since eqs.~(\ref{C4AQ}) and (\ref{C4QQ}), in contrast
to eq.~(\ref{S4mass}), now indeed involve all four vector fields. It turns
out that in dealing with this question, it is important {\em not to impose a
mixing scheme for the neutral vector fields} from the outside\cite{Vers}.
Once the possibilities of formation of vacuum expectation values (VEV's)
have been located by the charge-and-hypercharge diagonalization of the
preceding section, the present formalism will produce such a scheme, and
indeed a qualitatively correct picture of vector masses, all by itself.

         {\bf VEV formation}. In sect.~5 we pinpointed, on the basis of
minimality properties, the separable $\chi^a \, \chi^b$ term of
eq.~(\ref{Xicomp}) as the place where formation of vacuum-expectation values
should occur. In the present, four-fields situation, the natural extension
is to rely instead on the decomposition of eq.~(\ref{Xcomp}), generalized,
as in eq.~(\ref{XQbar}), to read 
\be
 {\bar Q} \lrp x \rrp \, \cdot \, {\bar Q}^{\dagger} \lrp x \rrp    \E  
{\bar P} \cdot {\bar P}^{\dagger} \, + \, {\bar q} \cdot 
                                        {\bar q}^{\dagger} \  , 
\label{Xbarcomp}
\ee
with obvious definitions. In the same vein,
\be
{\bar \phi}(x) \E \exp [ig_4\sigma h(x)] \, \phi (x)
\label{phibar}
\ee   
represents the ``hypercharged'', complex generalization of the
four-isovector (\ref{qtrans}). To find out which scalars can develop
nonzero VEV's, we should examine the eigenvectors -- in the present case,
the elements of the matrix $\bar \Phi$, and more specifically of the
four-isovector $\bar \phi$ -- of electric charge. Since the complex
dressing factor commutes, and the $C_4$ term in either form only involves
the real and symmetric combination
\be
\bar Q \, {\bar Q}^{\dagger} \E  Q \, Q^T \E X \  , 
\label{QQtee}
\ee
we may immediately take over the observations made on eq.~(\ref{qtrans}):
it is the component fields $\phi_2$ and $\phi_3$ that are electrically neutral
and therefore qualify for VEV formation. We combine this with the plausible
postulate that VEV's do not mix different eigenvalues of hypercharge -- in
other words, that they respect a kind of superselection rule with respect
to hypercharge. We first pursue the consequences of choosing $\phi_2$,
\be
\la 0 \lvl {\hat \phi}_2(x) \rvl 0 \ra  \E  \frac{1}{\sqrt{2}} \, v \, ,
\label{vev}
\ee
so that a four-dimensional extension of eq.~(\ref{vacexp}) holds:
\bea
\phi(x) \  &=& \  \frac{v}{\sqrt{2}} \begin{pmatrix}
                                         0                 \\
                                         1                 \\
                                         0                 \\
                                         0                  
                                     \end{pmatrix}            \quad  + \quad
                                     \begin{pmatrix}
                                         \phi_1 (x)        \\
                                         \phi_2^{\prime} (x) \\  
                                         \phi_3 (x)        \\
                                         \phi_4 (x)        
                                     \end{pmatrix}         \\     
&=& \  \frac{1}{\sqrt{2}} \, \mathbf{v} \  \qquad  + \qquad \phi^{\prime} (x) ,
\label{decvev}                               
\eea
where $\phi_4 (x) \E  \phi_1^{\star} (x)$. This, incidentally, is also
the isospin-hypercharge assignment of the VEV-forming neutral scalar in the
standard Higgs scenario. (The other possibility will be
commented upon below). The extraction of mass-squared terms from the $C_4$
action integrals will then proceed by performing the substitutions
\be
\lrp \Phi (x) \, \cdot \, \Phi^{\dagger} ( x) \rrp^{A\,B} \  \longrightarrow
\phi_A (x) \, \cdot \, \phi^{\dagger}_B(x)    \   \longrightarrow
\delta^{A\,2} \,\delta^{B\,2} \, \frac{1}{2} \, v^2
\label{subvev}
\ee
on one of the $QQ^T$ factors. We remark in passing that, strictly speaking,
it is only the {\em bilinear} correspondence of (\ref{subvev}), rather than
the linear one in (\ref{decvev}), that is necessary in order to exhibit the
VEV contributions to the action. In other words, we only need to assume that
the {\em two-point} Schwinger function of $\phi_2$ develop a constant condensate
$ \frac{1}{2} \, v^2 $, but not necessarily that this constant stems from
its disconnected term.

         {\bf Vector masses}. Start from the form (\ref{C4AQ}) of the
seagull term and relate the $\bar Q {\bar Q}^{\dagger}$ to $\Phi \Phi^{\dagger}$
by writing
\be
\lrp \bar Q {\bar Q}^{\dagger} \rrp^{A\,B}  \E  \lrp U^{\dagger} \rrp^{A\,C} \,
\lrp \Phi (x) \, \cdot \, \Phi^{\dagger} (x) \rrp^{C\,D} \,  U^{D\,B} \ ,
\label{QQTtrans}
\ee
then perform the substitution (\ref{subvev}), and read off the necessary
$U^{\dagger}$ and $U$ matrix elements from eqs.~(\ref{Umat}/\ref{Udagmat}).
You end up with the replacement
\be
\lrp \bar Q {\bar Q}^{\dagger} \rrp^{A\,B}  \longrightarrow \, \frac{1}{4} \, v^2
\lrp \delta^{\,A \,3} \, + \, \delta^{\,A \,4} \, \rrp \,
\lrp \delta^{\,B \,3} \, + \, \delta^{\,B \,4} \, \rrp \  .
\label{QQrep}
\ee
As for the matrix products $I_C^{\dagger} I_D$ figuring in both terms of the
$C_4$ action term (\ref{C4split}), we observe that in both cases they are
contracted with a quantity symmetric under exchange of indices $C$ and $D$.
We may therefore symmetrize those products as $\frac{1}{2} \, (I_C^{\dagger} I_D
+  I_D^{\dagger} I_C)$. The latter matrices are hermitean. Their imaginary
parts are therefore antisymmetric, and in a trace $tr_4$ with the symmetric
product (\ref{QQtee}) they make no contribution. Overall, the substitution
\bea
I_C^{\dagger} I_D  \quad \longrightarrow \quad M^2_{\,C\,D} \  ;
\label{IIsub}                                                     \\
M^2_{\,C\,D} \E \Re \lcp \frac{1}{2} \, (I_C^{\dagger} I_D + I_D^{\dagger}I_C) \rcp
\label{Mmat}
\eea
is permitted. Combining this with (\ref{QQrep}), we find that the ``seagull''
action (\ref{C4AQ}) splits off a mass term for the  vector fields,
\be
S_{\mathrm{vmass}} [ \, A \, ]   \E   \frac{1}{2} \, \int \! d^4 x \  
\lrp m^2 \rrp_{\, C,\, D}  \  A_{\mu}^C(x)   A_{\mu}^D(x) \  ,  
\label{vecmass}
\ee
where the mass-squared matrix $m^2$ is real and symmetric:
\be
\lrp m^{2} \rrp_{\, C, \, D}   \E  \frac{1}{2} \, g_2^{\,2} v_2 \,
\sum_{A,B=3}^4  \lrp M^2_{\,C\,D} \rrp^{A\,B} \  .
\label{vecmat}
\ee
Evaluating the matrices (\ref{Mmat}) from eq.~(\ref{Isimp}), we have for
their matrix elements,
\begin{align}           
\lrp M^2_{\,C\,D} \rrp^{A\,B}  \E  \frac{1}{4} \lsp \delta^{\,A \,B}\delta^{\,C \,D}
\, - \, \frac{1}{2} \lrp \delta^{\,A \,C}\delta^{\,B \,D}  \,+\,  \delta^{\,A \,D}
\delta^{\,C \,B} \rrp \rsp \,+\, \frac{3}{4} \lcp \delta^{\,A \,B}\,
(\delta^{\,C \,4} \delta^{\,D \,4}) \,+  \right. \qquad \qquad              \non
\left.          \delta^{\,C \,D}\,(\delta^{\,A \,4} \delta^{\,B \,4})
\, - \,  \frac{1}{2} \lsp \delta^{\,A \,C} \,
  (\delta^{\,B \,4} \delta^{\,D \,4}) \,+\, \delta^{\,A \,D}\,
  (\delta^{\,C \,4} \delta^{\,B \,4}) \,+\, \delta^{\,B \,C}\,
  (\delta^{\,A \,4} \delta^{\,D \,4}) \,+\, \delta^{\,B \,D}\,
  (\delta^{\,A \,4} \delta^{\,C \,4})    \rsp  \rcp \  .
\label{Melem}
\end{align}
Using these in eq.~(\ref{vecmat}), we obtain a vector mass-squared matrix
with block-diagonal structure,
\be
m^2  \E  \frac{1}{2} \, g_2^{\,2} \, v^2 \, \cdot
\begin{pmatrix}
  \frac{5}{4} \, \Id_2  &  {\textbf {\large 0}} \\
  {\textbf {\large 0}}  &       N_0^{\,2}
\end{pmatrix}                                            \ ,
\label{vecsqua}
\ee
where the matrix $N_0^{\,2}$ in the $A^3 \,-\, A^4$ subspace is given by
\be
N_0^{\,2}  \E   
\begin{pmatrix}
        1    &    -1       \\
       -1    &     1
\end{pmatrix}                                            \ .
\label{neutmat}
\ee
The latter matrix, visibly degenerate, has eigenvalues 2 and 0. The
corresponding eigenvectors give the $A^3 \,-\, A^4$ combinations that
diagonalize the quadratic form of eq.~(\ref{vecmass}),
\begin{align}
Z_{\mu} (x)  &=  \frac{1}{\sqrt{2}} \lsp A_{\mu}^3(x) \,-\,  A_{\mu}^4(x) \rsp \ ,
\label{zet}                                                           \\
A_{\mu} (x)  &=  \frac{1}{\sqrt{2}} \lsp A_{\mu}^3(x) \,+\,  A_{\mu}^4(x) \rsp \ , 
\label{photon}
\end{align}   
the latter being massless. The fully diagonalized form of $m^2$ thus reads,
\be
( m^2 )_{\text{diag}}    \E   g_2^{\,2} \, v^2 \, \cdot \text{diag}
\, \lcp \  \frac{5}{8}, \   \frac{5}{8}, \   1, \   0 \  \rcp \  .
\label{vecdiag}
\ee
On the other hand, in the mass-degenerate $A^1 \,-\, A^2$ subspace, where
any linear combination of these two fields is an eigenvector, the argument
follows the familiar pattern: the charge superselection rule singles out
the charge-conjugate pair
\begin{align}
W^{+} (x)  &=  \frac{1}{\sqrt{2}} \lsp A_{\mu}^1(x) \,-\,i\, A_{\mu}^2(x) \rsp \ ,
\label{wplus}                                                           \\
W^{-} (x)  &=  \frac{1}{\sqrt{2}} \lsp A_{\mu}^1(x) \,+\,i\, A_{\mu}^2(x) \rsp \ , 
\label{wminus}
\end{align}   
as the physical ones. What is noteworthy here is that once the modified
gauge-covariant derivative of eq.~(\ref{Atens}) has been established,
the mechanics of the $I_A$ matrices, when triggered by the VEV formation
(\ref{vev}) {\em but with no further input}, produces all by itself a
qualitatively correct picture of vector masses: a massive, charge-conjugate
pair, a neutral vector with slightly larger mass, and a massless photon.

The heavy-vector masses and their ratio are,
\be
m_W  \E  g_2 \, v \, \sqrt{\frac{5}{8}} ; \qquad m_Z  \E  g_2 \, v ;
\qquad m_W \, / \, m_Z  \E  \sqrt{\frac{5}{8}} \, \approx \, 0.79 \  .
\label{heavy}
\ee
The empirical value is $ \approx \, 0.87 $\cite{Part}. The comments made above
for eq.~(\ref{ratio}) apply again: there is no accurate fit with experiment,
but given the facts that these are tree-level values, and that they have been
obtained without adjustable parameters, the result (\ref{heavy})
would seem to be tolerable.

By comparison, we recall that in the standard Higgs scenario the counterparts
to eqs.~(\ref{vecsqua}/\ref{neutmat}) and eqs.~(\ref{zet}/\ref{photon})
are parameterized by the {\em electroweak mixing angle}, $\theta_W$, which    
is defined in analogy to eqs.~(\ref{cg}/\ref{sg}), but for the couplings
$g_2, \, g_1$
of the ordinary covariant derivative (\ref{oldcov}). It is a new constant of
nature, to be adjusted to experiment, and it appears (at tree level) in two
roles: first, it gives the $m_W \, / \, m_Z$ ratio as $c_W \E \cos (\theta_W)$,
from which it is determined empirically as $c_W \, \approx \, 0.87, \text{or} \,
\theta_W \, \approx \, 28.7 $ degrees. Second,
it appears in a degenerate hermitean matrix analogous to (\ref{neutmat}),
and consequently produces for the mixing scheme analogous to eqs.~(\ref{zet}
/, \ref{photon}) an $SO(2)$ orthogonal matrix parameterized by this very angle.
In the present formalism, these two roles are decoupled: the mass ratio
(again at tree level) is as in (\ref{heavy}), an algebraic number, while
the orthogonal mixing matrix has an angle of $45$ degrees.

It remains to comment on the alternative to eq.~(\ref{vev}), where
${\hat \phi}_3(x)$ rather than ${\hat \phi}_2(x)$ is allowed to develop
a vacuum expectation. One finds quickly that mass-squared eigenvalues
remain the same, and that the only change to eqs.~(\ref{zet}/\ref{photon})
is an exchange of the plus and minus signs, which amounts to redefining
$-A^4$ as $A^4$. Since $A^4$ appears only bilinearly in the original action,
this clearly changes nothing.

The end result of this section is then the desired form of the
vector-mass term (\ref{vecmass}) in the action,
\be
S_{vmass} [ \, W\,Z \, ]   \E   \frac{1}{2} \, \int \! d^4 x \  
\lsp  2 m_W^{\,2} \  W^{+}_{\mu}(x) \, W^{-}_{\mu}(x)
\, + \, m_Z^{\,2} \  Z_{\mu}(x) \, Z_{\mu}(x)  \  \rsp \  , 
\label{Svmass}
\ee
with $m_W$ and $m_Z$ as in eq.~(\ref{heavy}).

{\bf Scalar masses}. The discussion of sect.~5 on mass formation for the
scalars, based purely on the $\Xi$ fields, also needs to be reconsidered in
the new context created by the introduction of hypercharge. A convenient
starting point is eq.~(\ref{C4QQ}), with both the symmetrization (\ref{IIsub})
and the transform of eq.~(\ref{QQTtrans}) applied:
\begin{align}
\frac{1}{2} \,  C_4^{(Q)} \lsp \bar{Q}; \tilde{g} \sqrt{2} \rrp    \E
{\tilde g}^{\ 2} \, \int \! d^4 x \, \frac{1}{4} \, \lrp M^2_{\,C\,D} \rrp^{A\,B}
\qquad \qquad                                                           \non
\lsp U^{\dagger} \,
\lrp \Phi (x) \, \cdot \, \Phi^{\dagger} (x) \rrp \, U \rsp^{C,\,D}
\lsp U^{\dagger} \,
\lrp \Phi (x) \, \cdot \, \Phi^{\dagger} (x) \rrp \, U \rsp^{B,\,A} \  .
\label{C4sym}
\end{align}
Upon introducing the $U$ transform of the decomposition (\ref{Xbarcomp})
into (\ref{C4sym}), and taking eq.~(\ref{geescal}) into account, we have
\begin{align}
\frac{1}{2} \, C_4^{(Q)} \lsp \bar{Q}; \tilde{g} \sqrt{2} \rrp \E
(2g_2)^{\ 2} \, \int \! d^4 x \, \frac{1}{4} \, \lrp M^2_{\,C\,D} \rrp^{A\,B}
\qquad \qquad \qquad \qquad \qquad \qquad                              \non
\lcp \lrp U^{\dagger} \rrp^{C,\,E}  \lsp \lrp \Pi \cdot {\Pi}^{\dagger} \rrp^{E,\,F}
+ \phi_E  \cdot {\phi}^{\dagger}_{\,F} \rsp  U^{F,\,D} \rcp   
\lcp \lrp U^{\dagger} \rrp^{B,\,G}  \lsp \lrp \Pi \cdot {\Pi}^{\dagger} \rrp^{G,\,H}
+ \phi_G  \cdot {\phi}^{\dagger}_{\,H} \rsp  U^{H,\,A} \rcp \  ,
\label{C4decomp}
\end{align}
where $\Pi$ is the $U$ transform of the matrix $P$ of (\ref{Pmat}):
\be
\Pi (x)  \E  U \, P (x) \, U^{\dagger} \  .
\label{Pimat}
\ee
We now put to work three useful properties of the symmetrized 16-plet
(\ref{Mmat}) of matrices: 

\begin{itemize}
\item The contraction of $M^2$ with one and the same four-isovector
in all four indices vanishes. This is exemplified by the four-$\phi$s term
of (\ref{Mmat}): by eq.~(\ref{qtrans}),
\bea
\lrp M^2_{\,C\,D} \rrp^{A\,B}  \lcp \lrp U^{\dagger} \rrp^{C,\,E} \lsp \phi_E(x)\cdot
     {\phi}^{\dagger}_{\,F}(x)  \rsp \,  U^{F,\,D}   \  \lrp U^{\dagger} \rrp^{B,\,G}
     \lsp \phi_G(x)  \cdot {\phi}^{\dagger}_{\,H}(x) \rsp \, U^{H,\,A} \rcp   \non
     \E \lrp M^2_{\,C\,D} \rrp^{A\,B}\ q_C \,q_D \,q_B\,q_A
     \E  \, 0 \  . \qquad \qquad \qquad
     \label{qqqq}
\eea 
This implies, in particular, that {\em there is no $\phi$ mass term}, nor
are there cubic or quartic $\phi$ self-interactions. The
$\phi$ fields remain massless; they figure as quasi-Goldstone bosons.
(The property, actually, is already true for contraction with one
four-isovector in only three of the four indices).

\item Besides being symmetric, by construction, both in their index pair
$(A,B)$ and in the matrix-enumeration pair $(C,D)$, the sixteen matrices
possess the remarkable additional symmetry  
\be
\lrp M^2_{\,C\,D} \rrp^{\,A\,B}  \E  \lrp M^2_{\,A\,B} \rrp^{\,C\,D}    \  ,
\label{bigsym}
\ee
such that the ensemble may be viewed as a single supermatrix, with indices
consisting of symmetric pairs of ordinary matrix indices, and moreover
symmetric under the exchange of its two index pairs:
\be
{\hat M}^2_{\,A\,B , C\,D}  \E  {\hat M}^2_{\,C\,D , A\,B}  \  .
\label{supmat}
\ee
Since each of those pairs has ten different values, the supermatrix
${\hat M}^2$ is ten-dimensional, with $\frac{1}{2} \cdot 10 \cdot (10+1) \E 55$
different matrix elements. (It has some formal similarity
with a curvature tensor, except that the latter is {\em anti}symmetric
within each of its index pairs). The above statement is proven by
inspection of eq.~(\ref{Melem}).

Due to this additional symmetry, the two mixed $\Pi \, - \, 
\phi $ terms in eq.~(\ref{C4decomp}) are equal. In the final
form of (\ref{C4sym}),
\bea
\frac{1}{2} \,  C_4^{(Q)} \lsp \bar{Q}; \tilde{g} \sqrt{2} \rrp \E g_2^{\,2} \,
\int \! d^4 x \, \lrp M^2_{\,C\,D} \rrp^{A\,B} \qquad \qquad \qquad \qquad  \non
\lcp \lrp U^{\dagger} \rrp^{C,\,E}  \lsp 2 \,\phi_E \cdot {\phi}^{\dagger}_{\,F}
\,+\, \lrp \Pi \cdot {\Pi}^{\dagger} \rrp^{E,\,F} \rsp  U^{F,\,D} \rcp  
\lcp \lrp U^{\dagger} \rrp^{B,\,G}  \lsp \lrp \Pi \cdot {\Pi}^{\dagger} \rrp^{G,\,H}
\rsp  U^{H,\,A} \rcp \  , 
\label{C4Qsimp}
\eea
the mixed term therefore appears with a factor of $2$, whose origin is the
same as for the $2$ in the mixed $\Psi$-$\chi$ term of eq.~(\ref{masstr}),
and which will again be crucial for the scalars-to-vectors mass ratio.
In addition, there is of course a four-$\Pi$ self-interaction term,
which we do not discuss, except for pointing out that its coupling constant
is just the old $g_2^{\,2}$ of the nonabelian sector.
\item The supermatrix ${\hat M}^2$ is positive semi-definite on the subspace
of positive semi-definite, symmetric matrices. To check this, we
consider its mean value in an ordinary matrix $X \E X^T$ of this type:
\be
\lrp X \, , {\hat M}^2 \,X \rrp  \E \sum_{A,B}  \sum_{C,D} \, X_{A B} \,
{\hat M}^2_{A\,B , C\,D} \, X_{C D}   \  .
\label{Xmean}
\ee
Its evaluation gives
\bea
\lrp X \, , {\hat M}^2 \,X \rrp  &=&  \, \frac{1}{4} \lsp \lrp {\text{tr}}_4
\, X \rrp^2 \, - \, {\text{tr}}_4 \lrp X^2 \rrp \rsp                    \non
&+& \, \frac{3}{2} \lsp \lrp {\text{tr}}_4 \, X \rrp \, X_{4,4} \, - \,
\lrp X^2 \rrp_{4,4} \rsp  \  .
\label{posdef}
\eea
In the first line we recognize the four-dimensional analog of (\ref{trinq}),
which is positive semi-definite for the same reasons. But the second-line
term is also positive semi-definite: using the unitary matrix $V$ which
carries $X$ to its diagonal form $X_{\text{diag}} \E V \, X \, V^{\dagger}$,
with (by assumption) nonnegative eigenvalues $\lambda^A,\, A \E 1 \ldots 4$,
it becomes
\be
\frac{3}{2} \, \sum_{A\,B} \, | V^{4,\,A} |^2 \, \lambda^A \, \lambda^B
(\, 1 \,-\, \delta^{A\,B} \,) \  ,
\label{V4term}
\ee
from which, since $|V|^2$'s are between zero and unity, one concludes
\be
0 \, \le \, \frac{3}{2} \lsp .... \rsp \, \le \, \frac{3}{2} \lsp \lrp
{\text{tr}}_4 X \rrp^2 \, - \, {\text{tr}}_4 \lrp X^2 \rrp \rsp \  .
\label{V4estim}
\ee
Thus the integrand of $C_4^{\,(Q)}$ has a minimum of zero at $\Phi \Phi^{\dagger}
\E \phi \phi^{\dagger}$, i.\,e.\ at $\Pi \E 0$, a situation largely analogous to the
one in three isodimensions at the end of sect.~5. We can expect massive
Higgs-type particles in those $\Pi$ directions where there is a stable minimum,
but this time with more fields in the game.
\end{itemize}

We may now retrace the steps which in the vector case led from
eq.~(\ref{C4AQ}) to the isolation of the mass term (\ref{Svmass}), after the
replacement of eq.~(\ref{subvev}). We find that the mixed $\phi-\Pi$ part
of expression (\ref{C4Qsimp}) splits off a scalar-mass term,
\be
S_{\text{smass}} \lsp \Pi \rsp   \E  \frac{1}{2} \, \int \! d^4 x \    
\lrp \mu^2 \rrp_{\,A\,B} \, \lsp \Pi(x) \, \Pi^{\dagger}(x) \rsp^{\,B\,A}  \  ,    
\label{Ssmass}
\ee
where the scalar mass-squared matrix $\mu^2$ is given by
\be
\lrp \mu^2 \rrp_{\,A\,B}  \E  2 \lsp {\hat M}^2_{\,C\,D,\,E\,F} \, \frac{1}{2} \,
g_2^{\,2}\,v^2 \lrp \delta^{\,C\,3} \, + \, \delta^{\,C\,4} \, \rrp \,
             \lrp \delta^{\,D\,3} \, + \, \delta^{\,D\,4} \, \rrp \, \rsp
(\, U^{\dagger} \,)^{\,F\,B} \, U^{\,A\,E}    \  .
\label{scalmat}
\ee
In the square bracket, one recognizes the element $(m^2)_{\,E\,F}$ of the
vector mass-squared matrix (\ref{vecmat}). We therefore have the relation
\be
\mu^2  \E  2 \, U \, m^2 \, U^{\dagger}  
\label{scalvec}
\ee
between the two kinds of (squared) mass matrices. Upon carrying out the
$U$ transform one finds that {\em en passant} it diagonalizes the matrix
\bea
\mu^2 \  &=& \  g_2^{\,2}\,v^2  \, \text{diag} \lcp \, \frac{5}{4}, \  0, \  2,
\  \frac{5}{4} \, \rcp                                             \non
\qquad \  &=&  \text{diag} \lcp \, (m_W \sqrt{2})^2, \  0,  \
(m_Z \sqrt{2})^2,  \   (m_W \sqrt{2})^2 \, \rcp  \  .
\label{scaldiag}
\eea
The eigenvalues, apart from a different order of appearance, are twice
those for the squared vector masses, and $\sqrt{2}$ times those for the
vector masses themselves. Thus $T_3, \, Y=T_4$, and $\mu^2$ are all diagonal
in the same U(2) representation, as one would expect in an ordinary
Goldstone environment.

Since by eq.~(\ref{Pimat}) we have $\Pi\,\Pi^{\dagger} \E ( UP ) (UP)^{\dagger}$,
we may regroup the twelve $\Pi$ scalars as
\be
H^{A\,b}(x)  \E  \lsp U\, P(x) \rsp^{A\,b} \  , \qquad (\,b \E 1 \ldots 3\,) \ ,  
\label{Hafield}
\ee
for along with $P$, $UP$ also has vanishing fourth column. Then,
\be
\lsp \Pi(x) \,\Pi^{\dagger}(x) \rsp^{A\,B}  \E   H^{A\,c}(x) \,
\lrp H^{\dagger} \rrp^{c\,B} (x)  \  .
\label{PiPiHaHa}
\ee
By eq.~(\ref{scaldiag}), only $B=A$ elements appear in the action mass
term (\ref{Ssmass}), which therefore reads,
\be
S_{\text{smass}} \lsp \Pi \rsp   \E  \frac{1}{2} \, \int \! d^4 x \
\sum_{c=1}^3  \lcp (m_W \sqrt{2})^2 \, \lsp | H^{1\,c}(x) |^2 \,+\,
| H^{4\,c}(x) |^2 \rsp \ + \  (m_Z \sqrt{2})^2 \, | H^{3\,c}(x) |^2 \rcp \  .
\label{Ssdiag}
\ee
A glance at eq.~(\ref{Eldiag}) shows that $ H^{1\,c}(x) $ and $ H^{4\,c}(x) $
are charge conjugates, with electric charges $\pm \, 1$, while $ H^{3\,c}(x) $
is electrically neutral. Thus there are a mass-degenerate sextuplet of
charged Higgs-type scalars at a mass of $\mu^{(\pm)} \E m_W \sqrt{2}$, and a
degenerate triplet of neutral Higgs scalars at $\mu^{(0)} \E m_Z \sqrt{2}$.
Similarly to what we saw at the end of section 5, three other neutral
$H$ fields, $ H^{2\,c}(x) $, remain massless, as did all the $\phi$ or
$\phi^{\dagger}$ fields, although the former are not in the classical valley of
zero potential energy defined by $\Pi \E 0$. So there are now a total of
seven massless, candidate Goldstone fields.
\vspace{0.5cm}

\section{Loose Ends}
\setcounter{equation}{0}
%
We have seen that ``intrinsic'' scalar fields with several
desirable properties can be identified in an SU(2) $\otimes$ U(1) gauge
theory, and that one can go quite some distance in replaying with them
the standard scenario of spontaneous symmetry breaking and mass generation.
Nevertheless, the article, despite its length, has left several unresolved
questions. It is useful to enunciate these in closing.

\begin{itemize}
\item We have argued, in section 4, for the orthonormal vector system
$n^A_{\mu}$ being both $x$-independent and independent of gauge-field
configuration. A question that suggests itself, but to which at present
we have no clue, is whether that system can then be chosen arbitrarily,
or whether it retains some vestige of its gauge-theoretic construction
that sets it apart. It would be helpful here to have some explicit
examples of gauge fields for which that construction can be carried out
explicitly.

\item The true interrelationship of the seven massless fields, candidates
for Goldstone bosons, is arguably
not fully understood. The $\phi(x)$ doublet-plus-doublet of fields populate
the region $\Pi = 0$ of field space in which the quartic scalar potential
$C_4^{\,(Q)}$ assumes its minimum of zero, while remaining themselves massless.
Now the Higgs-type scalars of eq.~(\ref{Hafield}) also include a massless
triplet -- what precisely is the difference ? It may help to reflect on the
fact that {\em any} projector in isospace onto the direction of a single
isovector may become a zero minimum, namely if all fields {\em other than
this isovector} vanish. If we complement definition (\ref{Hafield}) by
\be
H^{A\,4}(x)  \E  \lsp U\, Q(x) \rsp^{A\,4} \E \phi_A(x) \  ,  
\label{Hafour}
\ee
then the projector onto the direction of the isovector $H^{2\,B},\,B=1
\ldots 4$ would also give a minimum of zero for $C_4^{\,(Q)}$, but in the region
where all fields $H^{A\,B}(x)$ with $A \neq 2$, including $\phi_A(x)$ with
$A \neq 2$, would vanish. Thus the two minima are in general incompatible --
in the density-matrix analogy invoked in section 5, they correspond to
projectors onto nonorthogonal pure states. The field configuration where
only $H^{2\,4}(x) \E \phi_2(x)$ is nonvanishing ``sits at the crossroads'';
it is the only one belonging to both classical minima, and in this sense
particularly susceptible to VEV formation. This may cast some light on the
special role of $\phi_2$ in our treatment, but it does not fully elucidate
the respective roles of the two remaining massless triplets in each minimum.

\item We have not touched upon the practically important problem of finding
a {\em unitarity gauge}, in which the massless scalars would be gotten rid
of at the expense of a loss of manifest renormalizability. For this problem,
the observation in section 7 of the presence of a six-parameter group of
motions in isospace may be of relevance, although in ways that have not been
demonstrated. Nor have we exhibited manifestly renormalizable gauges of the
$R_{\xi}$ type that would be welcome for calculations.
\end{itemize}
\newpage
\begin{appendix}
\section{Gauge and BRS variations}
\setcounter{equation}{0}
%
\hspace*{10mm} {\bf Gauge variations of $Q$ scalars}.
Under infinitesimal local gauge transformations characterized by gauge
functions $\delta \, \theta^a (x), \  a \E 1,\,2,\,3$, and
$\delta \, \theta^4 (x)$, the $A^a$ and $B$ gauge fields undergo the
well-known changes
\bea            
       \delta \, A_{\mu}^a (x)  & \E &  \lsp \delta^{a\,b} \,
       \partial_{\mu} \, - \, g_2 \, \epsilon^{a\,b\,c} \, A_{\mu}^c (x) \rsp 
       \, \delta \, \theta^b (x),                                    
\label{Avar}                                                         \\
       \delta \, B_{\mu} (x)  & \E & \lsp \partial_{\mu} \, 
       \delta \, \theta^4 (x) \rsp .                                 
\label{Bvar} 
\eea  
These formulas remain valid for BRS transformations when replacing the
infinitesimal gauge functions according to
\be            
\delta \, \theta^a (x) \, \longrightarrow \, \lambda \, c^a (x); \qquad
\delta \, \theta^4 (x) \, \longrightarrow \, \lambda \, c^4 (x) ,
\label{BRSvar} 
\ee  
with $\lambda$ a Grassmann-valued, constant parameter ($\lambda^2 \E 0$).
By virtue of the ``projection'' relation (\ref{QprimAn}) (with prime
dropped) and the property (\ref{ninv}) of the $n^A$ vector system, it becomes
straightforward to derive from these the gauge/BRS variations of the
$Q^{A\,B}$ scalar fields. We merely list the results: 
\be
\delta \, Q^{A\,B}  \E  \lrp \delta \, A_{\mu}^A \, \rrp \  n_{\mu}^B \  ,
\label{Qvar}
\ee
with special cases,
\bea
\delta \, \Psi^{a\,b}  & \E &  \lcp \lsp  \, \delta^{a\,d} \,
       \partial_{\mu} \, - \, g_2 \epsilon^{a\,d\,c} \, A_{\mu}^c (x) \rsp 
       \, \delta \, \theta^d (x) \rcp \, n_{\mu}^b \  ,               
\label{Psivar}                                                       \\
\delta \, \chi^{a}     & \E &  \lcp \lsp  \, \delta^{a\,d} \,
       \partial_{\mu} \, - \, g_2 \epsilon^{a\,d\,c} \, A_{\mu}^c (x) \rsp 
       \, \delta \, \theta^d (x) \rcp \, n_{\mu}^4 \  ,              
\label{chivar}
\eea
and moreover,
\bea
\delta \, \eta^{b}    & \E &  \lsp  \, \partial_{\mu} \,  
       \delta \, \theta^4 (x) \rsp \, n_{\mu}^b \  ,                   
\label{etavar}                                                        \\
\delta \, \psi       & \E &   \lsp  \, \partial_{\mu} \,  
       \delta \, \theta^4 (x) \rsp \,  n_{\mu}^4 \  .               
\label{psivar}  
\eea
When eqs.~(\ref{Psivar}) and (\ref{chivar})  are rewritten as  
\bea
\delta \, \Psi^{a\,b}  & \E &   i \, g_2 \, \lsp \, \delta \, \theta^c (x) \,
       f_c \rsp^{\,a\,d} \, \Psi^{\,d\,b}  \ + \  \partial_{\mu}
       \, \delta \, \theta^a (x) \, n_{\mu}^b \  ,                
\label{Psief}                                                        \\
\delta \, \chi^{a}    & \E &    i \, g_2 \lsp \, \delta \, \theta^c (x) \,
       f_c \rsp^{\,a\,d} \, \chi^{d}  \ \ \  + \  \partial_{\mu}
       \, \delta \, \theta^a (x) \, n_{\mu}^4 \  ,                
\label{chief}
\eea
in terms of the adjoint-generator matrices $f_c$ of (\ref{fc}), one
recognizes the first terms on the r.~h.\ sides as infinitesimal versions
of homogeneous gauge transformations $\Psi \longrightarrow \tilde{U}(x) \,
\Psi , \ \chi \longrightarrow \tilde{U}(x) \, \chi$ with matrix
$ \tilde{U}(x) \E \exp \lsp  i \, g_2 \, \delta \, \theta^c (x) \, f_c \rsp$.
The additional inhomogeneous or ``abelian'' terms, involving
$ \partial_{\mu} \, \delta \, \theta^a (x) $, that our scalars inherit
from their parent gauge fields are what sets them apart from the usual
Higgs scalars. In eqs.~(\ref{etavar}) and (\ref{psivar}), only these abelian
terms appear.                                                      \\[5mm]
\hspace*{10mm} {\bf Gauge variations of $\bar{Q}$ scalars}.
From the definition of $\bar{Q}$, eq.~(\ref{Qbar}), extension of the above
to the complex $\bar{Q}$ scalars is simple:
\be
\delta_{\theta} \, \bar{Q}  \E  \exp (i g_4 h(x) F_4) \, \lcp
\delta_{\theta} \, Q \, +  \,  i g_4 F_4 \,\delta \theta^4(x) \, Q \rcp  \  . 
\label{bardel}
\ee
The results can be collected in a compact form by writing
\be
\delta_{\theta} \, \bar{Q}^{A\,B}  \E  \lsp e^{[i2g_2h(x)]} \, \partial_{\mu} \,
\delta \theta^A(x) \rsp  n^B_{\,\mu} \,
+ \, i \lsp g_2 \delta \theta^d(x) F_d \, + \, 2g_2 \delta \theta^4(x) \Id_4 
\rsp^{A\,C} \, {\bar Q}^{C\,B} \  . 
\label{delbar}
\ee
It is understood that for $d \E 1,\,2,\,3$,
\be
( F_d )^{A\,C} \E 0, \quad \textnormal{if A or C}\, \E 4 \  .
\label{epsfour}
\ee
\vspace{0.5cm}
\section{Explicit Matrix Representations}
\setcounter{equation}{0}
%
The matrices $\tilde{I}_E$ of eq.~(\ref{Itildmat}), defining the interaction
part (second line of eq.~(\ref{Atens})) of the covariant derivative
$\nabla_{\mu} \bar{Q}$, are listed below:
\be
{\tilde I}_1  \E
\begin{pmatrix} 
   0   &   0   &   0   & -s_g\sigma \\
   0   &   0   & -ic_g &   0        \\
   0   & +ic_g &   0   &   0        \\ 
   0   &   0   &   0   &   0 
\end{pmatrix}  \  ; \qquad
{\tilde I}_2  \E
\begin{pmatrix} 
   0   &   0   & +ic_g &   0        \\
   0   &   0   &   0   & -s_g\sigma \\
-ic_g  &   0   &   0   &   0        \\ 
   0   &   0   &   0   &   0 
\end{pmatrix}  \  ;
\label{exmatItild12}
\ee
\be
{\tilde I}_3  \E
\begin{pmatrix} 
   0   & -ic_g &   0   &   0        \\
+ic_g  &   0   &   0   &   0        \\
   0   &   0   &   0   & -s_g\sigma \\ 
   0   &   0   &   0   &   0
\end{pmatrix}  \  ;  \qquad
{\tilde I}_4  \E
s_g\sigma \, \cdot \,
\begin{pmatrix} 
   1   &   0   &   0   &   0   \\
   0   &   1   &   0   &   0   \\
   0   &   0   &   1   &   0   \\ 
   0   &   0   &   0   &   0 
\end{pmatrix}  \  .  \qquad
\label{exmatItild34}
\ee
For the first three of these, the hermitean parts, candidates for a set
of four-dimensional isospin-SU(2) generators, are
\be
{\tilde R}_1  \E
\begin{pmatrix} 
            0           &   0   &   0   & -\frac{1}{2} s_g\sigma \\
            0           &   0   & -ic_g &             0          \\
            0           & +ic_g &   0   &             0          \\ 
-\frac{1}{2} s_g\sigma  &   0   &   0   &             0 
\end{pmatrix}  \  ; \qquad
{\tilde R}_2  \E
\begin{pmatrix} 
   0   &             0           & +ic_g &              0         \\
   0   &             0           &   0   & -\frac{1}{2} s_g\sigma \\
-ic_g  &             0           &   0   &   0                    \\ 
   0   & -\frac{1}{2} s_g\sigma  &   0   &   0 
\end{pmatrix}  \  ; \qquad
\label{exmatRtild12}
\ee
\be
{\tilde R}_3  \E
\begin{pmatrix} 
   0   & -ic_g &             0           &             0          \\
+ic_g  &   0   &             0           &             0          \\
   0   &   0   &             0           & -\frac{1}{2} s_g\sigma \\ 
   0   &   0   & -\frac{1}{2} s_g\sigma  &             0
\end{pmatrix}  \  \E  \  
\begin{pmatrix} 
  c_g \tau_2     &            {\mathbf 0}                         \\
  {\mathbf 0}   &  -\frac{1}{2} s_g\sigma \tau_1 
\end{pmatrix}  \  .
\label{exmatRtild3}
\ee
The last term uses the $2 \oplus 2$ block-diagonal form.  From these, one
may evaluate the commutator
\be
\lsp {\tilde R}_1 \, , \, {\tilde R}_2 \rsp  \E 
\begin{pmatrix} 
         0         &           c_g^2 + (\frac{1}{2}s_g\sigma)^2 &  0 &  0  \\
-c_g^2 - (\frac{1}{2}s_g\sigma)^2   &  0  &             0            &  0  \\
   0   &   0   &             0           &      -i\,c_gs_g\sigma           \\ 
   0   &   0   &     -i\,c_gs_g\sigma     &             0
\end{pmatrix}  \  .
\label{Rtildcomm}
\ee
This has the same pattern of nonzero entries as $i\,{\tilde R}_3$. It is
therefore proportional to $i\,{\tilde R}_3$ if the two ratios of corresponding
elements coincide:
\be
\frac{ c_g^2 + (\frac{1}{2}s_g\sigma)^2 }{ c_g }   \E k \  ;  \qquad
\frac{-i\,c_gs_g\sigma }{-\frac{i}{2} s_g\sigma }  \E k \  ;
\label{proport}
\ee
with $k$ a constant. These conditions give $k \E 2\,c_g$ and $s_g\sigma \E
2\,c_g$, and therefore the rescaling of eqs.~(\ref{geescal}/\ref{Imat})
and the condition (\ref{Rcomm}) for SU(2) generators. The rescaled,
absolutely fixed $R$ matrices then read,
\be
R_1  \E  \frac{1}{2} \  
\begin{pmatrix} 
            0   &   0   &   0   &  -1          \\
            0   &   0   &  -i   &   0          \\
            0   &  +i   &   0   &   0          \\ 
           -1   &   0   &   0   &   0 
\end{pmatrix}  \  ; \qquad
R_2  \E  \frac{1}{2} \  
\begin{pmatrix} 
   0   &   0   &  +i   &   0                   \\
   0   &   0   &   0   &  -1                   \\
  -i   &   0   &   0   &   0                   \\ 
   0   &  -1   &   0   &   0 
\end{pmatrix}  \  ; \qquad
\label{exmatR12}
\ee
\be
R_3  \E  \frac{1}{2}  
\begin{pmatrix} 
   0   &  -i    &   0   &   0       \\
 + i   &   0    &   0   &   0       \\
   0   &   0    &   0   &  -1       \\ 
   0   &   0    &  -1   &   0
\end{pmatrix}  \  \E  \frac{1}{2} \  
\begin{pmatrix} 
    \tau_2      &    {\mathbf 0}    \\
 {\mathbf 0}    &     -\tau_1)  
\end{pmatrix}  \  ;                 \\
\label{exmatR3}
\ee
\be
R_4  \E  \frac{1}{2} \
\begin{pmatrix} 
   0   &  -i    &   0   &   0       \\
 + i   &   0    &   0   &   0       \\
   0   &   0    &   0   &   1       \\ 
   0   &   0    &   1   &   0
\end{pmatrix}  \  \E  \frac{1}{2} \  
\begin{pmatrix} 
    \tau_2      &    {\mathbf 0}     \\
 {\mathbf 0}    &      \tau_1)  
\end{pmatrix}  \  .
\label{exmatR4}
\ee
We next list explicit representations of the $L_A$ matrices of 
eqs.~(\ref{Lmats}) and (\ref{L4}):
\be
L_1  \E  -\frac{1}{2} \  
\begin{pmatrix} 
            0   &   0   &   0   &   1          \\
            0   &   0   &   0   &   0          \\
            0   &   0   &   0   &   0          \\ 
           -1   &   0   &   0   &   0 
\end{pmatrix}  \  ; \qquad
L_2  \E  -\frac{1}{2} \  
\begin{pmatrix} 
   0   &   0   &   0   &   0                   \\
   0   &   0   &   0   &   1                   \\
   0   &   0   &   0   &   0                   \\ 
   0   &  -1   &   0   &   0 
\end{pmatrix}  \  ; \qquad
\label{exmatL12}
\ee
\be
L_3  \E  -\frac{1}{2} \  
\begin{pmatrix} 
   0   &   0   &   0   &   0                   \\
   0   &   0   &   0   &   0                   \\
   0   &   0   &   0   &   1                   \\ 
   0   &   0   &  -1   &   0
\end{pmatrix}  \  \E  -\frac{1}{2} \  
\begin{pmatrix} 
  {\mathbf 0}   &  {\mathbf 0}                  \\
  {\mathbf 0}   &  i \, \tau_2  
\end{pmatrix}  \  ;                             \\
\label{exmatL3}
\ee
\be
L_4  \E \      
\begin{pmatrix} 
   1            &  \frac{i}{2}  &        0       &        0       \\
 -\frac{i}{2}   &       1       &        0       &        0       \\
       0        &       0       &        1       &  -\frac{1}{2}  \\ 
       0        &       0       &  -\frac{1}{2}  &        0
\end{pmatrix}  \  \E  
\begin{pmatrix} 
  \Id_2-\frac{1}{2}\tau_2   &           {\mathbf 0}                \\
       {\mathbf 0}          &  \frac{1}{2}(\Id_2+\tau_3-\tau_1)  
\end{pmatrix}  \  .                             
\label{exmatL4}
\ee
For the $T_A$ matrices in the charge-diagonal representation, the forms
of eqs.~(\ref{Tmat}/\ref{T4diag}) are sufficiently explicit. We
therefore list explicit forms only for the transforms $S_A$ of the $L_A$'s:
\be
S_1  \E  \frac{1}{4}  \  
\begin{pmatrix} 
   0    &   1   &  -1   &   0             \\
  -1    &   0   &   0   &   1             \\
   1    &   0   &   0   &  -1             \\ 
   0    &  -1   &   1   &   0 
\end{pmatrix}  \  \E  -S_1^T \  ;   \qquad
S_2  \E  \frac{1}{4}  \  
\begin{pmatrix} 
   0   &   -i   &  +i   &    0            \\
  -i   &    0   &   0   &   -i            \\
   i   &    0   &   0   &    i            \\ 
   0   &   -i   &  +i   &    0 
\end{pmatrix}  \  \E  -S_2^{\dagger} \  ;   \qquad
\label{exmatS12}
\ee
\be
S_3  \E  \frac{1}{4}  \  
\begin{pmatrix} 
   0   &   0    &   0   &   0             \\
   0   &   0    &   2   &   0             \\
   0   &  -2    &   0   &   0             \\ 
   0   &   0    &   0   &   0
\end{pmatrix}  \  \E  \frac{1}{4} \  
\begin{pmatrix} 
     {\mathbf 0}       &  \tau_1 - i\,\tau_2  \\
 -\tau_1 - i\,\tau_2   &   {\mathbf 0}  
\end{pmatrix}  \  \E  -S_3^T;              \\
\label{exmatS3}
\ee
\be
S_4  \E  \frac{1}{2} \
\begin{pmatrix} 
   1   &   0    &   0   &   0              \\
   0   &   0    &   1   &   0              \\
   0   &   1    &   2   &   0              \\ 
   0   &   0    &   0   &   3
\end{pmatrix}  \  \E  \frac{1}{4} \  
\begin{pmatrix} 
   \Id_2 + \tau_3     &    \tau_1 - i\,\tau_2  \\
 \tau_1 + i\,\tau_2   &       5\Id_2 - \tau_3 
\end{pmatrix}  \  .
\label{exmatS4}
\ee
One verifies that $T_4\,+\,S_4$ gives the $U$ transform of $I_4$,
\be
U \,I_4 \, U^{\dagger}  \E  \frac{1}{2} \
\begin{pmatrix} 
   1   &        0        &        0        &    0          \\
   0   &   \frac{1}{2}   &   \frac{1}{2}   &    0          \\
   0   &   \frac{1}{2}   &   \frac{1}{2}   &    0          \\ 
   0   &        0        &        0        &    1
\end{pmatrix}  \  .  
\label{UI4U}
\ee
Finally we write the explicit form of the transformed matrix $\Phi \E
U\,Q\,U^{\dagger}$ mentioned in section 9:
\be
\Phi(x) \E  U \  Q(x) \  U^{\dagger}  \E                    
\begin{pmatrix} 
\Phi_{1\,1}   &  -\Phi^{\,*}_{4\,2} &  -\Phi^{\,*}_{4\,3}  &   \Phi^{\,*}_{4\,1}    \\
\Phi_{2\,1}   &     \Phi_{2\,2}    &     \Phi_{2\,3}     &  -\Phi^{\,*}_{2\,1}    \\
\Phi_{3\,1}   &     \Phi_{3\,2}    &     \Phi_{3\,3}     &  -\Phi^{\,*}_{3\,1}    \\ 
\Phi_{4\,1}   &     \Phi_{4\,2}    &     \Phi_{4\,3}     &   \Phi^{\,*}_{1\,1}
\end{pmatrix}  \  .  
\label{exmatPhi}
\ee
Here the sixteen real fields of eq.~(\ref{Qpart}) appear recombined into
six complex and four real amplitudes:
\bea
\Phi_{1\,1}\ &=&\ \frac{1}{2} \lsp\lrp\quad\Psi_{1\,1} + \Psi_{2\,2} \rrp
                           +\,i\,\lrp   \Psi_{1\,2} - \Psi_{2\,1} \rrp \rsp , \\
\Phi_{4\,1}\ &=&\ \frac{1}{2} \lsp \lrp-\,\Psi_{1\,1} + \Psi_{2\,2} \rrp
                           -\,i\,\lrp   \Psi_{1\,2} + \Psi_{2\,1} \rrp \rsp , \\
\vspace{1mm}
\Phi_{2\,1}\ &=&\ \frac{1}{2} \lsp-\,\lrp \Psi_{3\,1} + \eta_1     \rrp
                           -\,i\,\lrp   \Psi_{3\,2} + \eta_2     \rrp \rsp , \\
\Phi_{3\,1}\ &=&\ \frac{1}{2} \lsp-\,\lrp \Psi_{3\,1} - \eta_1     \rrp
                           -\,i\,\lrp   \Psi_{3\,2} - \eta_2     \rrp \rsp , \\
\vspace{1mm}
\Phi_{4\,2}\ &=&\ \frac{1}{2} \lsp \lrp \ \Psi_{1\,3} + \chi_1     \rrp
                           +\,i\,\lrp   \Psi_{2\,3} + \chi_2     \rrp \rsp , \\
\Phi_{4\,3}\ &=&\ \frac{1}{2} \lsp \lrp \ \Psi_{1\,3} - \chi_1     \rrp
                           +\,i\,\lrp   \Psi_{2\,3} - \chi_2     \rrp \rsp ,
\eea
\vspace{1mm}                                                               \\
\bea
\Phi_{2\,2}\ &=&\ \frac{1}{2} \lsp  \lrp \Psi_{3\,3} + \eta_3  \rrp
                             + \, \lrp   \chi_3    + \psi    \rrp \rsp , \\
\Phi_{2\,3}\ &=&\ \frac{1}{2} \lsp  \lrp \Psi_{3\,3} + \eta_3  \rrp
                             - \, \lrp   \chi_3    + \psi    \rrp \rsp , \\
\Phi_{3\,2}\ &=&\ \frac{1}{2} \lsp  \lrp \Psi_{3\,3} - \eta_3  \rrp
                             + \, \lrp   \chi_3   - \psi    \rrp \rsp ,  \\
\Phi_{3\,3}\ &=&\ \frac{1}{2} \lsp  \lrp \Psi_{3\,3} - \eta_3  \rrp
                             - \, \lrp   \chi_3    - \psi    \rrp \rsp .
\eea
The corresponding matrix elements for the Matrix $\Pi(x)$ of eq.~(\ref{Pimat})
are obtained for $\chi_a \E 0$ and $\psi \E 0$. One then notes that the
two middle columns of this matrix become equal, which checks with our earlier
observation that $\det P \E 0$.
\end{appendix}
\newpage


\begin{thebibliography}{99}
\bibitem{Vers} Version 1 of this article, in its section 9, imposed the
               standard neutral mixing, parameterized by the electroweak
               mixing angle $\theta_W$, from the outset, and consequently
               was forced into a complicated yet unsatisfactory construction
               for vector-mass formation, which the present version shows
               to be unnecessary.
\end{thebibliography}
\end{document}